\documentclass[journal]{IEEEtran}
\usepackage{amsmath,amsfonts}
\usepackage[ruled,commentsnumbered]{algorithm2e}
\usepackage{array}
\usepackage[caption=false,font=normalsize,labelfont=sf,textfont=sf]{subfig}
\usepackage{textcomp}
\usepackage{stfloats}
\usepackage{url}
\usepackage{verbatim}
\usepackage{graphicx}
\usepackage{cite}
\usepackage{doi}
\usepackage{multirow}
\usepackage{adjustbox}
\usepackage{enumitem}
\usepackage{color}

\begin{document}
	\title{A Model Fusion Distributed Kalman Filter\\ for Non-Gaussian Measurement Noise}
	
	\author{Xuemei Mao, Jiacheng He, Gang Wang, Bei Peng, Kun Zhang, Song Gao, Jian Chen
		% <-this % stops a space
		\thanks{Manuscript received xxx; accepted x May xxx. This article was recommended for publication by Associate Editor xxx and Editor xxx upon evaluation of the reviewers’ comments. (Corresponding author: Bei Peng.)}
		\thanks{Xuemei Mao, Bei Peng, Jiacheng He, Kun Zhang, Song Gao, and Jian Chen are with the School of Mechanical and Electrical Engineering, University of Electronic Science and Technology of China, Chengdu, 611731, China (e-mail:beipeng@uestc.edu.cn).}% <-this % stops a space
		\thanks{Gang Wang is with the School of Information and Communication Engineering, University of Electronic Science and Technology of China, Chengdu, 611731, China.}
		\thanks{Copyright (c) 20xx IEEE. Personal use of this material is permitted. However, permission to use this material for any other purposes must be obtained from the IEEE by sending a request to pubs-permissions@ieee.org.}}
	
	\markboth{IEEE Internet of Things Journal}%
	{Shell \MakeLowercase{\textit{et al.}}: A Model Fusion Distributed Kalman Filter for Non-Gaussian Measurement Noise}
	
	% % column for its text to clear the IEEEpubid mark.
	
	\maketitle
	
	\begin{abstract}
		Wireless sensor networks (WSNs) represent a critical research domain within the Internet of Things (IoT) technology. The distributed Kalman filter (DKF) has garnered significant attention as an information fusion method for WSNs. However, effectively handling non-Gaussian environments remains a crucial challenge for DKF. This paper proposes a solution by partitioning the noise distribution into multiple Gaussian components, thereby approximating the measurement model with sub-models. We introduce a model fusion distributed Kalman filter (MFDKF) that combines sub-models by assuming independent random processes for the model's transition probabilities. The expectation maximization (EM) algorithm is employed to estimate the relevant parameters. To address specific requirements in WSNs that demand high consensus or have limited communication, two derivative algorithms, namely consensus MFDKF (C-MFDKF) and simplified MFDKF (S-MFDKF), are proposed based on consensus theory. The convergence of MFDKF and its derivative algorithms is analyzed. A series of simulations demonstrate the effectiveness of MFDKF and its derivative algorithms.
	\end{abstract}
	
	
	\section{Introduction}
	\IEEEPARstart{T}{he} wireless sensor networks (WSNs) constitute a pivotal area of research within the framework of Internet of Things (IoT) technology, contributing significantly to intelligent systems, including identification, positioning, tracking, and supervision \cite{ma2022method, zhang2022consensus, liu2020multistep, talebi2016distributed}. Fusion of multiple sensor data for state estimation is a fundamental function of WSNs that has attracted considerable attention due to its improved performance and practicality \cite{YANG20142070,he2020distributed, sun2017multi}. 
	
	Centralized Kalman filter (KF) is a straightforward fusion architecture that collects data from all sensor nodes into a central node \cite{sun2004multi}. The advantage of centralized KF is its ability to utilize all available information to achieve the best estimation. Still, its computationally intensive and communication-heavy requirements pose challenges for scalability, limiting its expansion \cite{modalavalasa2021review}. Additionally, the failure of the central node can lead to a complete breakdown of estimations, indicating poor robustness. To address these issues, a distributed KF (DKF) is proposed for WSNs \cite{olfati2005distributed}. The distributed architecture employs independent local KF (LKF) at each node and utilizes point-to-point communication with neighboring nodes exclusively. By leveraging local information sharing, DKF reduces reliance on a central node and lowers communication demands, thereby enhancing the system's scalability, robustness, and flexibility. Yet, DKF prohibits information exchange between non-adjacent nodes, which increases the risk of estimation inconsistencies among nodes, posing challenges for coordinated tasks \cite{olfati2007consensus}. The consensus algorithm has been effectively integrated into the DKF, reducing disagreement among nodes in WSNs \cite{olfati2007distributed}. Reference \cite{olfati2009kalman} designates this class of algorithms as the Kalman-Consensus Filter and provides theoretical support for its optimality, stability, and performance.
	
	Much of the aforementioned research is conducted under the assumption that sensor measurement noise follows a Gaussian distribution. In practical scenarios, measurement noise may exhibit non-Gaussian characteristics due to complex environments, disturbances, and other uncertainties \cite{he2022generalized, wang2022centralized, he2024gaussian}. For instance, the measurement noise from acoustic sensors in underwater environments exhibits a heavy-tailed characteristic due to factors such as complex propagation media, multipath transmission, environmental disturbances, and nonlinear effects \cite{9167469}. The non-Gaussian characteristics of noise in practical applications contradict the Gaussian assumptions made in theory, resulting in covariance matrices that fail to accurately represent the actual noise properties \cite{fan2022background}. This discrepancy introduces bias in the data fusion process of the DKF, ultimately affecting the accuracy of state estimation.
	
	Consequently, the study of DKF  in non-Gaussian noise environments has become a prominent research area, as state estimation in WSNs is a critical foundation for the IoT \cite{modalavalasa2021review}. Distributed Particle Filtering (DPF) can effectively handle nonlinear and non-Gaussian systems, which has led to the emergence of many DPF methods that incorporate consensus algorithms \cite{coates2004distributed, yu2016distributed, li2017distributed}. These methods introduce a consensus mechanism to facilitate information sharing and fusion in multi-agent systems, thereby improving the accuracy and robustness of state estimation. However, the main challenge of DPF lies in the heavy computational burden caused by a large number of particles. Despite various improvements, DPF still exhibits substantial computational intensity compared to other parametric algorithms \cite{coates2004distributed}. The distributed Student's t filter (DTF) \cite{xu2018distributed} provides a solution with reduced computational complexity while performing well in non-Gaussian scenarios by assuming the noise is heavy-tailed and modeling it as a Student's t distribution. Another form of the DTF based on information filtering has been derived, enabling suboptimal distributed fusion through parameter approximation to reduce computational complexity \cite{yan2020distributed}. Additionally, the DTF based on the suboptimal arithmetic average fusion method is specifically designed to address the issue of unknown correlations among sensors \cite{li2022multi}. Nevertheless, these algorithms depend on the choice of the prior distribution, and they perform well only when the selected Student's t distribution closely matches the actual noise characteristics \cite{8214971}.
	
	Recently, information-theoretic learning (ITL) methods have been introduced for state estimation in non-Gaussian environments \cite{chen2017maximum,chen2019minimum, wang2021numerically, he2023generalized}. These methods optimize state estimates by minimizing output entropy, effectively addressing non-Gaussian noise without the limitations of selecting a model that aligns with the actual noise distribution. When the measurement noise is non-Gaussian, the ITL-based DKF outperforms the conventional DKF (CDKF) \cite{he2020distributed} as it can capture higher-order moment information of the error and adjust the estimation accordingly. The distributed maximum correntropy KF (DMCKF) is formulated in \cite{wang2019distributed} by introducing a new gain matrix and leveraging the maximum correntropy criterion (MCC) \cite{liu2007correntropy}. Wang et al. also propose a novel DMCKF based on the consensus average method extending its application to non-linear systems \cite{wang2021distributed}. To enhance the communication efficiency, a covariance intersection DMCKF is proposed, using a matrix weight obtained by MCC \cite{hu2022efficient}. In ITL, the minimum error entropy (MEE) \cite{zhang2015convergence} criterion considers the Euclidean distance between a zero-mean Gaussian distribution and the actual error probability density function, making it preferable to MCC when handling complex noise \cite{8267224}. The distributed minimum error entropy KF (DMEEKF) is proposed based on the MEE criterion \cite{feng2023distributed} to enhance the robustness in non-Gaussian environments. 
	
	To better elucidate the operational mechanisms of DMCKF and DMEEKF, we categorize the noises in non-Gaussian environments into two types: one type consists of small noises that predominantly occur and have a small covariance, denoted as $\boldsymbol{R_b}$; the other type is impulsive noise, which appears with low probability and has a larger covariance, denoted as $\boldsymbol{R_u}$. The noise covariance $\boldsymbol{R}$ used for state estimation is derived from noise sampling and its  diagonal elements lie between $\boldsymbol{R_b}$ and $\boldsymbol{R_u}$, which prevents it from accurately reflecting the true characteristics of the noise, thereby limiting the performance of CDKF in non-Gaussian noise environments.
	 
	The effectiveness of both DMCKF and DMEEKF in non-Gaussian noise arises from the minimization of output entropy, which introduces a tuning mechanism to address the occurrence of impulsive noise \cite{fan2022background}. When impulsive noise is present, the noise covariance $\boldsymbol{R}$ in state estimation is appropriately increased through a tuning matrix, allowing it to better reflect the current true noise characteristics and enhance estimation performance. When impulsive noise is absent, the tuning matrix approximates the identity matrix. Thus, it can be observed that the performance enhancement of DMCKF and DMEEKF primarily stems from their effective capture of impulsive noise, while in most cases, small noise is approximated by $\boldsymbol{R}$, which still remains greater than the actual $\boldsymbol{R_b}$. This insight provides us with an approach to improving the DKF in non-Gaussian noise environments, specifically by finding ways to obtain covariance data that aligns with the current noise distribution.  
	
	The above research motivates us to model non-Gaussian noise using a Gaussian Mixture Model (GMM) \cite{reynolds2009gaussian}, which decomposes the noise into multiple Gaussian components with varying parameters. Our goal is to identify the covariance that best matches the current actual noise distribution through these Gaussian components, thereby enhancing the DKF's state estimation performance. In WSNs, each sensor node simultaneously experiences noise interference, with the measurement noise decomposed into multiple Gaussian components. This results in numerous potential combinations of noise models for the LKF. 
	\begin{figure}[!ht]
		\centering
		\includegraphics[width=0.95\linewidth]{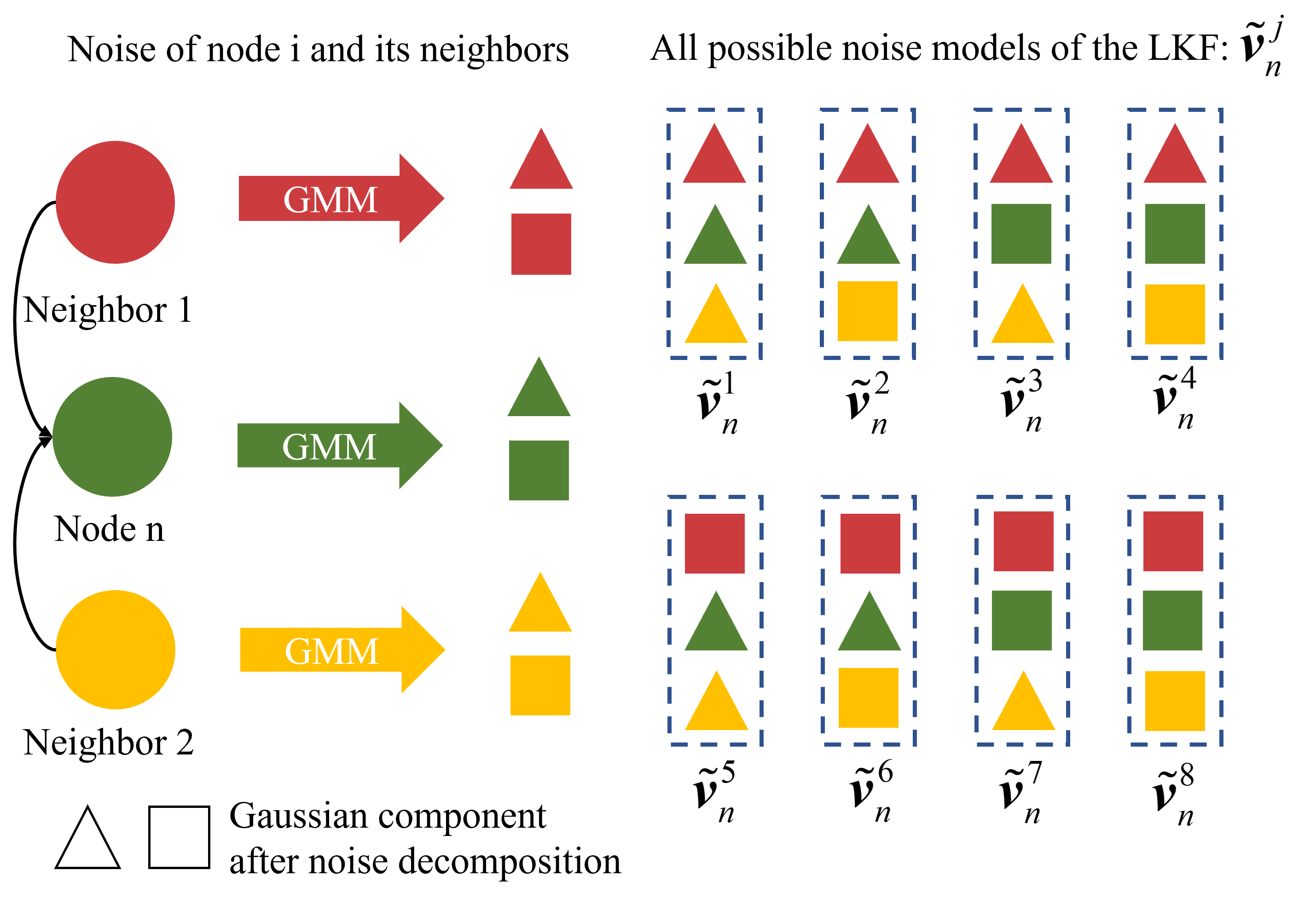}
		\caption{Noise Decomposition and Measurement Modeling}
		\label{mao1}
	\end{figure}
	
	Fig. \ref{mao1} illustrates an example of the measurement fusion based on the Gaussian components. Node $n$ is connected to two neighboring nodes, which contribute their measurements for state estimation using the LKF. Initially, the measurement noise at each node is decomposed into two Gaussian components using the GMM framework. The noise models of node $n$ and its neighboring nodes are then fused to form the measurement noise model for the LKF. Notably, this decomposition process yields eight distinct instances representing possible occurrences of measurement noise within the LKF at any given moment. Each instance serves as a measurement model, and the LKF is applied individually to each model to derive the corresponding state estimates. Finally, these state estimates are fused based on the likelihood of each observation model \cite{fan2022background, fan2021interacting} to enhance the performance of the DKF in non-Gaussian environments. 
	
	Using multiple Gaussian components enables a more precise representation of the diverse noise characteristics encountered by the WSNs. This study provides the following contributions:
	\begin{enumerate}
		\item We propose a model fusion DKF (MFDKF) tailored to non-Gaussian environments. MFDKF divides sub-models with distinct fusion noises and fuses their estimates to improve state estimation in the presence of non-Gaussian noise.
		\item To address the specific requirements of certain WSNs, we introduce two derived algorithms of MFDKF based on consensus theory: consensus MFDKF (C-MFDKF) for enhanced node consensus and simplified MFDKF (S-MFDKF) for reduced communication and computation. 
		\item Some analyses are given to illuminate the performance of the proposed MFDKF and its derived algorithms, including mean error analysis, mean-square error analysis, and computational complexity. Simulations are performed to indicate the effectiveness of the MFDKF and its derivative algorithms.  
	\end{enumerate}
	
	The remainder of this paper is structured as follows. Section \ref{pre} gives some preparatory knowledge of MFDKF and explains the design motivation. Section \ref{Propose MFDKF} covers the implementation of the MFDKF algorithm. Two derived algorithms of MFDKF, C-MFDKF and S-MFDKF are presented in Section \ref{derived MFDKF}. The performance analysis of the algorithms is presented in Section \ref{performance}. Simulations and conclusions are discussed in Section \ref{simulations} and Section \ref{conclusions}.
	
	\textit{Notations:} The notation ${\left(  \cdot  \right)^T}$ represents the transpose operation, while $diag(\cdot)$ creates a block diagonal matrix from the given elements. $\left| \cdot \right|$ is the operator that calculates the determinant of a matrix, and $E$ denotes the expectation operator. $O(.)$ is the same term infinitesimal of the variable. Additionally, $\otimes$ represents the Kronecker product, and $vec$ reshapes a matrix into a column vector by stacking its columns.
	
	\section{Preliminaries} \label{pre}
	\subsection{Analysis of LKF}
	A target measured by a WSN can be formulated as follows:
	\begin{align} 
		{{\boldsymbol{x}}_k} = {{\boldsymbol{A}}_{k - 1}}{{\boldsymbol{x}}_{k - 1}} + {{\boldsymbol{w}}_{k - 1}}, \label{process_fun_1} \\
		{{\boldsymbol{z}}_{n,k}} = {{\boldsymbol{H}}_{n,k}}{{\boldsymbol{x}}_k} + {{\boldsymbol{v}}_{n,k}}.
	\end{align}
	Here, ${{\boldsymbol{x}}_k} \in {\mathbb{R}^{p \times 1}}$ is the unknown state vector at time $k$, ${{\boldsymbol{z}}_{n,k}} \in {\mathbb{R}^{q \times 1}}$ is the measurement vector of node $n$, which is available in the time $k$. The known ${{\boldsymbol{A}}_k} \in {\mathbb{R}^{p \times p}}$ and ${{\boldsymbol{H}}_{n,k}} \in {\mathbb{R}^{q \times p}}$ respectively represent the state transition matrix and the measurement matrix of node $n$. ${{\boldsymbol{w}}_k} \in {\mathbb{R}^{p \times 1}}$ and ${{\boldsymbol{v}}_{n,k}} \in {\mathbb{R}^{q \times 1}}$ refer to system noise and measurement noise, respectively. Supposing they are uncorrelated noises, their corresponding covariance matrices are ${{\boldsymbol{Q}}_{\boldsymbol{k}}}$ and ${{{\boldsymbol{R}}_{{\boldsymbol{n,k}}}}}$, respectively.
	
	The WSN consisting of $N$ nodes can be represented by an undirected graph ${\cal G} = \left( {{\cal V},{\cal E}} \right)$, where ${\cal V} = \left\{ {1,2, \cdots ,N} \right\}$ denotes the set of nodes, and ${\cal E} = {\cal V} \times {\cal V}$ represents the edge set that defines the neighboring relationships among the nodes. The neighbors of a node $n \in {\cal V}$ are denoted as ${{\cal A}_n} = \left\{ {m|\left( {m,n} \right) \in {\cal E}} \right\}$, and self-edge is permitted. The number of elements in ${{\cal A}_n}$ is denoted as $d_n$.
	
	The purpose of DKF is to incorporate information from neighboring nodes. Define the fusion matrices as 
	\begin{equation}
			\begin{array}{*{20}{l}}
				{{{{\boldsymbol{\tilde z}}}_{n,k}} = {{\left[ {{\boldsymbol{z}}_{{n_1},k}^T,{\boldsymbol{z}}_{{n_2},k}^T, \cdots ,{\boldsymbol{z}}_{{n_{{d_n}}},k}^T} \right]}^T} \in {\mathbb{R}^{q{d_n} \times 1}},}\\
				{{{{\boldsymbol{\tilde H}}}_{n,k}} = {{\left[ {{\boldsymbol{H}}_{{n_1},k}^T,{\boldsymbol{H}}_{{n_2},k}^T, \cdots ,{\boldsymbol{H}}_{{n_{{d_n}}},k}^T} \right]}^T} \in {\mathbb{R}^{q{d_n} \times p}},}\\
				{{{{\boldsymbol{\tilde v}}}_{n,k}} = {{\left[ {{\boldsymbol{v}}_{{n_1},k}^T,{\boldsymbol{v}}_{{n_2},k}^T, \cdots ,{\boldsymbol{v}}_{{n_{{d_n}}},k}^T} \right]}^T} \in {\mathbb{R}^{q{d_n} \times 1}},}
			\end{array}
	\end{equation}
	where, ${n_1},{n_2}, \cdots ,{n_{{d_n}}} \in {{\cal A}_n}$.
	Then, the fusion LKF measurement model for node $n$ can be described as
	\begin{equation} \label{central_obser}
		{{\boldsymbol{\tilde z}}_{n,k}} = {{\boldsymbol{\tilde H}}_{n,k}}{{\boldsymbol{x}}_k} + {{\boldsymbol{\tilde v}}_{n,k}},
	\end{equation}
	the covariance matrix of ${{\boldsymbol{\tilde v}}_{n,k}}$ is obtained by  
	\begin{equation}
			{{\boldsymbol{\tilde R}}_{n,k}} = diag\left( {{{\boldsymbol{R}}_{{n_1},k}},{{\boldsymbol{R}}_{{n_2},k}}, \cdots ,{{\boldsymbol{R}}_{{n_{{d_n}}},k}}} \right)  \in {\mathbb{R}^{q{d_n} \times q{d_n}}}.
	\end{equation}
	Applying the KF to model (\ref{process_fun_1}) and (\ref{central_obser}), the iterations of LKF in CDKF \cite{he2020distributed} are summarized as
	\begin{align}
		{{{\boldsymbol{\bar x}}}_{n,k}} = {{\boldsymbol{A}}_{k - 1}}{{{\boldsymbol{\hat x}}}_{n,k - 1}}, \label{CDKF_prediction}\\ 
		{{{\boldsymbol{\bar P}}}_{n,k}} = {{\boldsymbol{A}}_{k - 1}}{{{\boldsymbol{\hat P}}}_{n,k - 1}}{\boldsymbol{A}}_{k - 1}^T + {{\boldsymbol{Q}}_{k - 1}}, \label{CDKF_P_}\\
		{{\boldsymbol{S}}_{n,k}} = {{{\boldsymbol{\tilde H}}}_{n,k}}{{{\boldsymbol{\bar P}}}_{n,k}}{\boldsymbol{\tilde H}}_{n,k}^T + {{{\boldsymbol{\tilde R}}}_{n,k}}, \label{S_c}\\
		{{\boldsymbol{K}}_{n,k}} = {{{\boldsymbol{\bar P}}}_{n,k}}{\boldsymbol{\tilde H}}_{n,k}^T{\boldsymbol{S}}_{n,k}^{ - 1}, \label{K_c}\\
		{{{\boldsymbol{\bar v}}}_{n,k}} = {{{\boldsymbol{\tilde z}}}_{n,k}} - {{{\boldsymbol{\tilde H}}}_{n,k}}{{{\boldsymbol{\bar x}}}_{n,k}}, \label{v_c}\\
		{{{\boldsymbol{\hat x}}}_{n,k}} = {{{\boldsymbol{\bar x}}}_{n,k}} + {{\boldsymbol{K}}_{n,k}}{{{\boldsymbol{\bar v}}}_{n,k}}, \label{xk}\\
		{{{\boldsymbol{\hat P}}}_{n,k}} = \left( {{\boldsymbol{I}} - {{\boldsymbol{K}}_{n,k}}{{{\boldsymbol{\tilde H}}}_{n,k}}} \right){{{\boldsymbol{\bar P}}}_{n,k}}, \label{CDKF Mk}
	\end{align}
	where ${{{\boldsymbol{\hat P}}}_{n,k}}$ is the error covariance matrix of state estimation ${{{\boldsymbol{\hat x}}}_{n,k}}$ and ${{{\boldsymbol{\bar P}}}_{n,k}}$ is the error covariance matrix of one step prediction ${{{\boldsymbol{\bar x}}}_{n,k}}$.
	
	Utilizing (\ref{S_c}), (\ref{K_c}) and the matrix inversion lemma \cite{9661360}, the state update (\ref{xk}) can be rewritten as
	\begin{align}
		{{{\boldsymbol{\hat x}}}_{n,k}} = {{\boldsymbol{\Pi }}_{n,k}}{{{\boldsymbol{\bar x}}}_{n,k}} + {{\boldsymbol{\Sigma }}_{n,k}}{{{\boldsymbol{\tilde z}}}_{n,k}}, \label{xk1} \\
		{{\boldsymbol{\Pi }}_{n,k}} = {{{\boldsymbol{\hat P}}}_{n,k}}{\boldsymbol{\bar P}}_{n,k}^{ - 1}, \label{alpha} \\
		{{\boldsymbol{\Sigma }}_{n,k}} = {{{\boldsymbol{\hat P}}}_{n,k}}{\boldsymbol{\tilde H}}_{n,k}^T{\boldsymbol{\tilde R}}_{n,k}^{ - 1}. \label{beta}
	\end{align}
	Equation (\ref{xk1}) shows that the estimation ${{{\boldsymbol{\hat x}}}_{n,k}}$ can be simplified as a combination of the one-step prediction ${{{\boldsymbol{\bar x}}}_{n,k}}$ and the fusion measurement ${{{\boldsymbol{\tilde z}}}_{n,k}}$. Equations (\ref{alpha}) and (\ref{beta}) show that the combination coefficient ${{\boldsymbol{\Pi }}_{n,k}}$ and ${{\boldsymbol{\Sigma }}_{n,k}}$ depend on the error covariance matrix of one step prediction ${\boldsymbol{\bar P}}_{n,k}^{{ - 1}}$ and the covariance matrix ${{\boldsymbol{\tilde R}}_{n,k}}$. The analysis above reveals the principles of DKF. Increased noise leads to less accurate measurements, necessitating a greater consideration of predicted values in state estimation to reduce the impact of measurements. In CDKF, the accuracy of state estimation is compromised due to the unchanging ${{{\boldsymbol{R}}_{{n,k}}}}$, which deviates significantly from the true value in the presence of impulsive noise.
	
	Consider the mixed Gaussian noise as an example, which follows the distribution $v \sim 0.9 {\cal N}\left( {0 ,0.01^2} \right) + 0.1 {\cal N}\left( 0 ,100^2 \right)$. The statistic value of the covariance, ${{{\boldsymbol{R}}_{{n,k}}}}$, is approximately ${10^3}$, lying between $0.01^2$ and $100^2$. In the absence of impulsive noise, the measurement is reliable, and a large value should be assigned to the matrix ${{\boldsymbol{\Sigma }}_{n,k}}$. Nevertheless, a relatively smaller value is utilized due to the relatively large ${{{\boldsymbol{R}}_{{n,k}}}}$. In the presence of impulsive noise, the measurement becomes inaccurate, and the weight in the matrix ${{\boldsymbol{\Sigma }}_{n,k}}$ should decrease. But, it still retains a relatively larger value due to the unchanged ${{{\boldsymbol{R}}_{{n,k}}}}$, which is smaller than the actual value. In summary, in non-Gaussian scenarios, the actual noise characteristics may deviate significantly from the modeled Gaussian distribution. Maintaining the original ${{\boldsymbol{R}}_{n,k}}$ means the filter might not correctly adjust the weights ${{\boldsymbol{\Sigma }}_{n,k}}$, leading to suboptimal performance.
	
	The DMCKF \cite{wang2019distributed} and DMEEKF \cite{feng2023distributed} utilize the error between measurement and predicted measurement error ${{{\boldsymbol{\bar v}}}_{n,k}}$ (the calculation method refers to (\ref{v_c})) to detect the presence of impulsive noise and adjust the weights accordingly. When impulsive noise occurs, it correspondingly increases the noise covariance used to calculate ${{\boldsymbol{\Sigma }}_{n,k}}$, thereby reducing the weights in the matrix ${{\boldsymbol{\Sigma }}_{n,k}}$. In the absence of impulsive noise, the noise covariance to calculate ${{\boldsymbol{\Sigma }}_{n,k}}$ is close to the sampled value. A problem arises in the absence of impulsive noise, the weights in ${{\boldsymbol{\Sigma }}_{n,k}}$ are smaller than intended, as the sampled noise covariance is larger than the actual value due to the occasionally impulsive noise. This motivates us to approximate non-Gaussian noise as a combination of multiple Gaussian components, facilitating the ${{{\boldsymbol{R}}_{{n,k}}}}$ used to calculate ${{\boldsymbol{\Sigma }}_{n,k}}$ to align to the true value in non-Gaussian environments.
	
	\subsection{Gaussian Mixture Model} \label{EM}
	The GMM is a linear combination of Gaussian noise, and its probability density function \cite{reynolds2009gaussian} is given by: 
	\begin{align}
		p({{\boldsymbol{v}}_k}) = \sum\nolimits_{i = 1}^\kappa  {{\gamma ^i}{\cal N}({\boldsymbol{\mu }}_k^i,{\boldsymbol{R}}_k^i)} ,\sum\nolimits_{i = 1}^\kappa  {{\gamma ^i}}  = 1,\\
		{\cal N}({\boldsymbol{\mu }}_k^i,{\boldsymbol{R}}_k^i) = \frac{{\exp \left[ {\frac{1}{2}{{({{\boldsymbol{v}}_k} - {\boldsymbol{\mu }}_k^i)}^T}{{({\boldsymbol{R}}_k^i)}^{ - 1}}({{\boldsymbol{v}}_k} - {\boldsymbol{\mu }}_k^i)} \right]}}{{{{(2\pi )}^{q/2}}|{\boldsymbol{R}}_k^i{|^{1/2}}}},
	\end{align}
	where ${{{\boldsymbol{v}}_{k}}}$ is a $q$ dimensional noise vector, ${\boldsymbol{\mu }}_k^i$ and ${\boldsymbol{R}}_k^i$ severally denote the mean and covariance matrix of the $i$-th component. $\gamma ^i$ is the weight of the component $i$.
	
	Given enough noise samples, the parameters of each Gaussian component can be obtained using the expectation maximization (EM) algorithm \cite{min2019robust, guo2017augmented}. Denoting the number of components of node $n$ as $\kappa_n$, the Gaussian components of the ${{\boldsymbol{v}}_{n,k}}$ obtained for node $n$ can be denoted as
	\begin{equation}
		\left\{ {{\boldsymbol{v}}_{n,k}^1,{\boldsymbol{v}}_{n,k}^2, \cdots ,{\boldsymbol{v}}_{n,k}^{{\kappa _n}}} \right\},
	\end{equation}
	where, ${\boldsymbol{v}}_{n,k}^{{\kappa _n}}$ represents the $\kappa_n$-th component of the noise ${{\boldsymbol{v}}_{n,k}}$.
	The parameters of the components obtained from the EM algorithm are denoted as 
	\begin{equation}
			\begin{array}{*{20}{l}}
					\left\{ {{\boldsymbol{R}}_{n,k}^1,{\boldsymbol{R}}_{n,k}^2, \cdots ,{\boldsymbol{R}}_{n,k}^{{\kappa _n}}} \right\},
					\left\{ {{\boldsymbol{\mu }}_{n,k}^1,{\boldsymbol{\mu }}_{n,k}^2, \cdots ,{\boldsymbol{\mu }}_{n,k}^{{\kappa _n}}} \right\},\\
				{\left\{ {\gamma _n^1,\gamma _n^2,...,\gamma _n^{{\kappa _n}}} \right\},\sum\limits_{i = 1}^{{\kappa _n}} {\gamma _n^i}  = 1.}
			\end{array}
	\end{equation}
	Here, ${{\boldsymbol{R}}_{n,k}^{{\kappa _n}}}$, ${{\boldsymbol{\mu }}_{n,k}^{{\kappa _n}}}$, and ${\gamma _n^{{\kappa _n}}}$ represent the covariance matrix, mean vector, and weight of the Gaussian component ${{\boldsymbol{v}}_{n,k}^{{\kappa _n}}}$, respectively.
	\section{Proposed Model Fusion Distributed Kalman Filter} \label{Propose MFDKF}
	In this section, we initially approximate the ${{\boldsymbol{v}}_{n,k}}$ of each node using a GMM as described in Section \ref{EM}. The measurement model of node $n$ in the DKF can be approximated as a fusion of multiple sub-models as illustrated in Fig. \ref{mao1}. 
	It is important to note that the Gaussian components do not always have zero mean, which may cause the LKF to fail as a least mean square estimator. Inspired by the literature \cite{9655313}, which equates colored noise to Gaussian white noise through matrix augmentation, we propose to set the mean of the noise to a known systematic deviation. This allows us to focus on the remaining error after subtracting the mean, ensuring the error meets the zero mean condition. Finally, an MFDKF is obtained by fusing sub-estimations provided by the sub-models. To address anomalies in the application of the fusion method, an algorithm modification is implemented.
	
	By dividing the noise of each node into $\kappa_n$ Gaussian components (${\kappa _n} \ge 2)$, we can partition the measurement model of each node into several measurement sub-models. Let ${\boldsymbol{\tilde z}}_{n,k}^{j}$ represents a possible model of measurement for the LKF of node $n$, which can be viewed as the ${j}$-th submodel of the measurement. The LKF measurement model of node $n$ is represented as
		\begin{equation} \label{fusion measurement}
			{\boldsymbol{\tilde z}}_{n,k}^j = {{{\boldsymbol{\tilde H}}}_{n,k}}{{\boldsymbol{x}}_k} + {\boldsymbol{\tilde \mu }}_{n,k}^j + {\boldsymbol{\tilde v}}_{n,k}^j,
		\end{equation}
	where, 
	\begin{small}
			\begin{equation}
				\begin{array}{*{20}{l}}
					{{\boldsymbol{\tilde z}}_{n,k}^j = {{\left[ {{{({\boldsymbol{z}}_{{n_1},k}^{{j_{{n_1}}}})}^T}\!\!,{{({\boldsymbol{z}}_{{n_2},k}^{{j_{{n_2}}}})}^T}\!\!, \cdots \!\!,{{({\boldsymbol{z}}_{{n_{{d_n}}},k}^{{j_{{n_{{d_n}}}}}})}^T}} \right]}^T} \!\!\!\in {\mathbb{R}^{q{d_n} \times 1}},}\\					
					{{\boldsymbol{\tilde \mu }}_{n,k}^j = {{\left[ {{{({\boldsymbol{\mu }}_{{n_1},k}^{{j_{{n_1}}}})}^T}\!\!,{{({\boldsymbol{\mu }}_{{n_2},k}^{{j_{{n_2}}}})}^T}\!\!, \cdots \!,{{({\boldsymbol{\mu }}_{{n_{{d_n}}},k}^{{j_{{n_{{d_n}}}}}})}^T})} \right]}^T} \!\!\!\in {\mathbb{R}^{q{d_n} \times 1}}.}\\
					{{\boldsymbol{\tilde v}}_{n,k}^j = {{\left[ {\begin{array}{*{20}{l}}
										{{{({\boldsymbol{v}}_{{n_1},k}^{{j_{{n_1}}}} - {\boldsymbol{\mu }}_{{n_1},k}^{{j_{{n_1}}}})}^T},}\\
										{{{({\boldsymbol{v}}_{{n_2},k}^{{j_{{n_2}}}} - {\boldsymbol{\mu }}_{{n_2},k}^{{j_{{n_2}}}})}^T},\cdots,}\\
										{{{({\boldsymbol{v}}_{{n_{{d_n}}},k}^{{j_{{n_{{d_n}}}}}} - {\boldsymbol{\mu }}_{{n_{{d_n}}},k}^{{j_{{n_{{d_n}}}}}})}^T}}
								\end{array}} \right]}^T} \in {\mathbb{R}^{q{d_n} \times 1}}.}
				\end{array}
			\end{equation}
	\end{small}
	${j_{{n_t}}} \in \left\{ {1,2, \cdots ,{\kappa _{{n_t}}}|{n_t} \in {{\cal A}_n}} \right\}$, ${\boldsymbol{v}}_{{n_t},k}^{{j_{{n_t}}}}$ represents the ${j_{{n_t}}}$-th Gaussian component of the noise ${{\boldsymbol{v}}_{{n_t},k}}$, whose covariance matrix, mean vector, and weight are ${\boldsymbol{R}}_{{n_t},k}^{{j_{{n_t}}}}$, ${\boldsymbol{\mu }}_{{n_t},k}^{{j_{{n_t}}}}$ and $\gamma _{{n_t}}^{{j_{{n_t}}}}$, respectively. The main distinction between ${\boldsymbol{\tilde z}}_{n,k}^{j}$ depends on the noise model ${\boldsymbol{\tilde v}}_{n,k}^j$. Fig. \ref{mao1} offers an example to obtain the possible noise model of the LKF measurement.
	The covariance matrix of ${\boldsymbol{\tilde v}}_{n,k}^{j}$ is denoted as
	\begin{equation}
			{\boldsymbol{\tilde R}}_{n,k}^j = diag({\boldsymbol{R}}_{{n_1},k}^{{j_{{n_1}}}},{\boldsymbol{R}}_{n2,k}^{{j_{{n_2}}}}, \cdots ,{\boldsymbol{R}}_{{n_{{d_n}}},k}^{{j_{{n_{{d_n}}}}}}) \in {\mathbb{R}^{q{d_n} \times q{d_n}}}.
		\end{equation}
	 Based on the decomposition, ${\boldsymbol{z}}_{{n_t},k}^{{j_{{n_t}}}}$ represents the ${j_{{n_t}}}$-th component of ${{\boldsymbol{z}}_{n_t,k}}$. 
	
	The quantity $L_n$ represents the number of LKF measurement sub-models associated with node $n$,  
	\begin{equation} \label{model number}
		{L_n} = \prod\limits_{{n_t} \in {{\cal A}_n}} {{\kappa _{{n_t}}}} .
	\end{equation}
	The probabilities of the measurement sub-models are calculated by
	\begin{equation} \label{calculate possibilities}
		\alpha _n^j = \prod\limits_{{n_t} \in {{\cal A}_n}} {\gamma _{{n_t}}^{{j_{{n_t}}}}} ,j \in \left\{ {1,2, \cdots ,{L_n}} \right\}.
	\end{equation}
	To compute the probability distribution of each sub-model at each time step, we assume that the model's transition probabilities follow an independent random process, with the transition probability defined as:
		\begin{equation}
			\tilde P_n^{ij} = \alpha _n^j,
		\end{equation}
		where $\widetilde P_n^{ij}$ is the transition probability of node $n$ from sub-model $i$ to sub-model $j$.
	
	The parameters of the model fusion are determined. The Interacting Multiple Model (IMM) \cite{xie2019adaptive} algorithm is typically employed in scenarios where system dynamics can switch between different modes or models. In this study, we utilize the core framework of the IMM algorithm to address the issue of switching between different observation models, thereby adapting to non-Gaussian noise environments. By facilitating interaction among different models, we can better manage variations in measurement noise and optimize the final state estimation results. 
	\begin{figure}[!ht]
		\centering
		\includegraphics[width=1\linewidth]{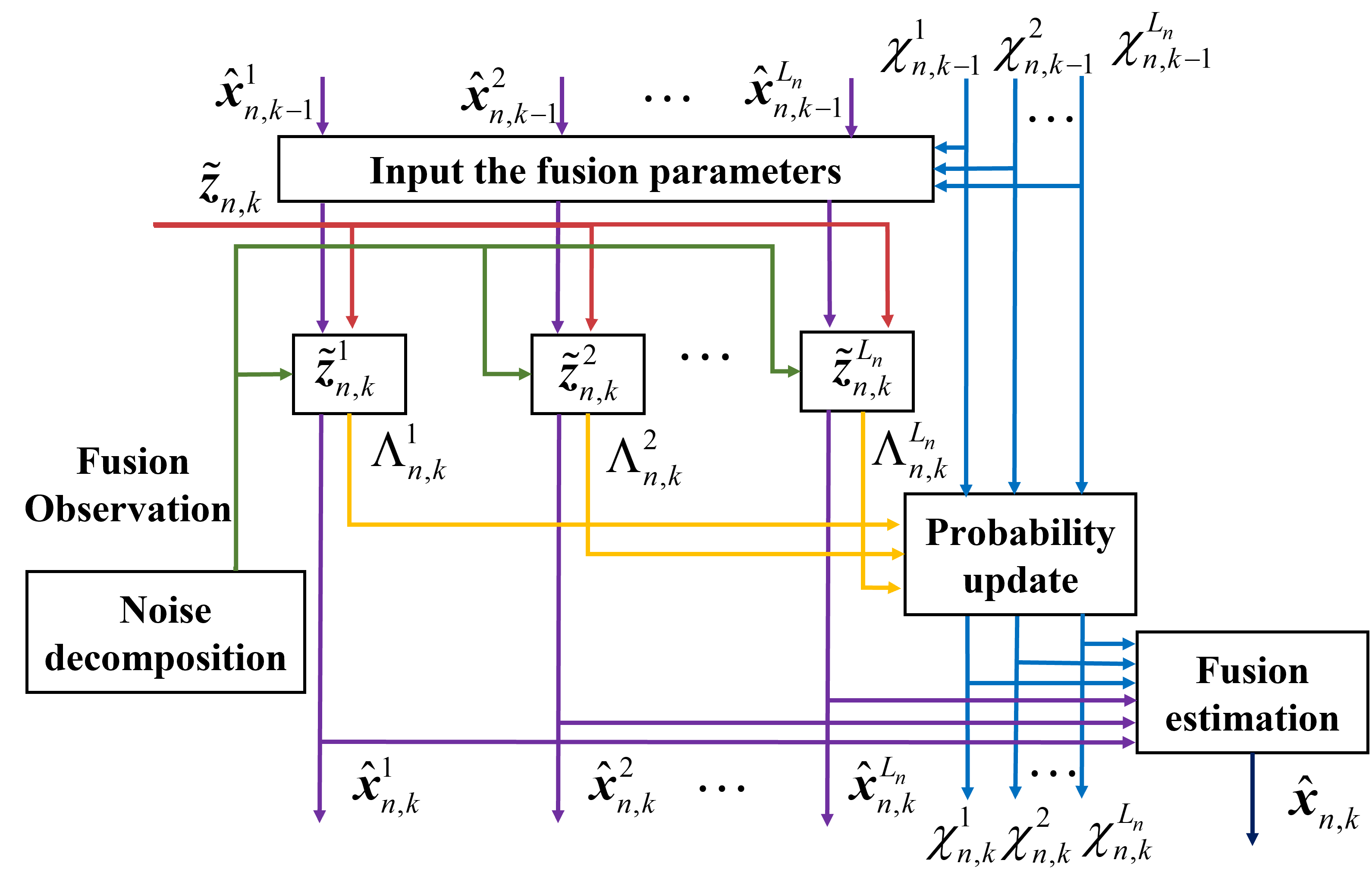}
		\caption{Process of the MFDKF}
		\label{mao2}
	\end{figure}
	
	To estimate the current state, separate estimations are performed for each possible model. These estimations are subsequently combined based on the probability of model occurrence, yielding the final estimation. The process is illustrated in Fig. \ref{mao2}. After adjusting the update of the model probability, the steps of MFDKF at each node are summarized in Algorithm \ref{MFDKF}, where $\chi _{n,k}^i$ and ${\boldsymbol{\hat x}}_{n,k}^{i}$ are the probability and estimation of the sub-model $i$ at the instant $k$, respectively. ${\boldsymbol{\hat P}}_{n,k}^{i}$ is the error covariance matrix of ${\boldsymbol{\hat x}}_{n,k}^{i}$.
	\begin{algorithm}[ht]
		\caption{Model Fusion Distributed Kalman Filter}
		\label{MFDKF} 
		Initialization: 
		$\chi _{n,0}^i = \alpha _n^i,{\boldsymbol{\hat x}}_{n,0}^i = {{{\boldsymbol{\hat x}}}_0},{\boldsymbol{\hat P}}_{n,0}^i = {{{\boldsymbol{\hat P}}}_0}$\;
		
		\For{$k=1:T$}{
			\For{$i=1:L_n$}{\For{$j=1:L_n$}{\begin{equation} \label{fusion possibility}
							\begin{array}{l}
								\pi _{n,k - 1}^{ij} = \tilde P_n^{ij}\chi _{n,k - 1}^i/\bar c_{n,k - 1}^j,\\
								\bar c_{n,k - 1}^j = \sum\nolimits_{i = 1}^{{L_n}} {\tilde P_n^{ij}\chi _{n,k - 1}^i} ;
							\end{array}
			\end{equation}}}
			\For{$j=1:L_n$}{
				\begin{equation} \label{state_fusion_func}
					{\boldsymbol{\hat x}}_{n,k - 1}^{pre\_j} = \sum\nolimits_{i = 1}^{{L_n}} {{\boldsymbol{\hat x}}_{n,k - 1}^i\pi _{n,k - 1}^{ij}} ;
				\end{equation}
				\begin{equation}\label{cov_xpre}
					\begin{array}{*{20}{l}}
						{{\boldsymbol{\hat P}}_{n,k - 1}^{pre\_j} = \sum\nolimits_{i = 1}^{{L_n}} {\pi _{n,k - 1}^{ij}} }\\
						{\left[ {\begin{array}{*{20}{l}}
									{{\boldsymbol{\hat P}}_{n,k - 1}^i + ({\boldsymbol{\hat x}}_{n,k - 1}^i - {\boldsymbol{\hat x}}_{n,k - 1}^{pre\_j})}\\
									{{{({\boldsymbol{\hat x}}_{n,k - 1}^i - {\boldsymbol{\hat x}}_{n,k - 1}^{pre\_j})}^T}}
							\end{array}} \right];}
					\end{array}
				\end{equation}
				The LKF for each sub-model:
				\begin{equation} \label{updating_funcj}
						\begin{array}{*{20}{l}}
							{{\boldsymbol{\hat x}}_{n,k}^j = {\boldsymbol{\bar x}}_{n,k}^j + {\boldsymbol{K}}_{n,k}^j{\boldsymbol{\bar v}}_{n,k}^j,}\\
							{{\boldsymbol{\bar x}}_{n,k}^j = {{\boldsymbol{A}}_{k - 1}}{\boldsymbol{\hat x}}_{n,k - 1}^{pre\_j},}\\
							{{\boldsymbol{\bar v}}_{n,k}^j = {{{\boldsymbol{\tilde z}}}_{n,k}} - {{{\boldsymbol{\tilde H}}}_{n,k}}{\boldsymbol{\bar x}}_{n,k}^j - {\boldsymbol{\tilde \mu }}_{n,k}^j,}\\
							{{\boldsymbol{\bar P}}_{n,k}^j = {{\boldsymbol{A}}_{k - 1}}{\boldsymbol{\hat P}}_{n,k - 1}^{pre\_j}{\boldsymbol{A}}_{k - 1}^T + {{\boldsymbol{Q}}_{n,k}},}\\
							{{\boldsymbol{S}}_{n,k}^j = {{{\boldsymbol{\tilde H}}}_{n,k}}{\boldsymbol{\bar P}}_{n,k}^j{\boldsymbol{\tilde H}}_{n,k}^T + {\boldsymbol{\tilde R}}_{n,k}^j,}\\
							{{\boldsymbol{K}}_{n,k}^j = {\boldsymbol{\bar P}}_{n,k}^j{\boldsymbol{\tilde H}}_{n,k}^T{{\left( {{\boldsymbol{S}}_{n,k}^j} \right)}^{ - 1}};}
						\end{array}
				\end{equation}
				\begin{equation} \label{p_hat_j}
					{\boldsymbol{\hat P}}_{n,k}^j = [{\boldsymbol{I}} - {\boldsymbol{K}}_{n,k}^j{{{\boldsymbol{\tilde H}}}_{n,k}}]{\boldsymbol{\bar P}}_{n,k}^j;
				\end{equation}
				\begin{equation} \label{likelihood}
					\Lambda _{n,k}^j = \frac{{\exp [ - \frac{1}{2}{{({\boldsymbol{\bar v}}_{n,k}^j)}^T}{{({\boldsymbol{S}}_{n,k}^j)}^{ - 1}}{\boldsymbol{\bar v}}_{n,k}^j]}}{{{{(2\pi )}^{q{d_n}/2}}|{\boldsymbol{S}}_{n,k}^j{|^{1/2}}}};
				\end{equation}
				\eIf{$\sum\nolimits_{j = 1}^{{L_n}} {\Lambda _{n,k}^j} \sim0$}{
					\begin{equation} \label{modified}
						\begin{array}{l}
							j = {{\mathord{\buildrel{\lower3pt\hbox{$\scriptscriptstyle\frown$}} 
										\over j} }_{\max \left( {\left| {{\boldsymbol{\tilde R}}_{n,k}^j} \right|} \right)}},\chi _{n,k}^{j = \mathord{\buildrel{\lower3pt\hbox{$\scriptscriptstyle\frown$}} 
									\over j} } = 1,\\
							\chi _{n,k}^{j \ne \mathord{\buildrel{\lower3pt\hbox{$\scriptscriptstyle\frown$}} 
									\over j} } = 0,{\boldsymbol{\tilde R}}_{n,k}^j = {\boldsymbol{\bar v}}_{n,k}^j{\left( {{\boldsymbol{\bar v}}_{n,k}^j} \right)^T};
						\end{array}
					\end{equation}
				}{\begin{equation} \label{proba_update}
							\chi _{n,k}^j = \bar c_{n,k}^j\Lambda _{n,k}^j/\sum\nolimits_{i = 1}^{{L_n}} {\bar c_{n,k}^j\Lambda _{n,k}^i} ;
				\end{equation}}
			}
			\begin{equation} \label{state fusion}
				{{{\boldsymbol{\hat x}}}_{{\boldsymbol{n}},{\boldsymbol{k}}}} = \sum\nolimits_{j = 1}^{{L_n}} {\chi _{n,k}^j{\boldsymbol{\hat x}}_{n,k}^j} ;
			\end{equation}
		}
	\end{algorithm}
	
	The update of the model probability in Algorithm \ref{MFDKF} is modified using (\ref{modified}). Direct application of the IMM framework is not feasible due to the likelihood function calculated from (\ref{likelihood}) potentially approaching zero when a majority of neighboring nodes experience strong impulsive noise. Consequently, the denominator in (\ref{proba_update}) tends to zero, resulting in the iteration collapse. Therefore, it is necessary to address the update of the model probability. One possible solution is to set $\chi _{n,k}^j = \bar c_n^j$. Regrettably, this approximation introduces significant instability to the algorithm.
	
	Another approximation is applied. As described in (\ref{modified}), neighboring nodes are assumed to be affected by the highest levels of impulsive noise, when  (\ref{likelihood}) approaches zero. We use the predicted measurement noise ${\boldsymbol{\bar v}}_{n,k}^j$ to obtain a more precise fusion covariance matrix ${\boldsymbol{\tilde R}}_{n,k}^j$. The simulation results in Section \ref{simulations} demonstrate that this method can operate smoothly.  
	
	\section{C-MFDKF and S-MFDKF Based on Consensus Fusion} \label{derived MFDKF}
	The MFDKF prohibits the exchange of information between non-neighboring nodes. There is no centralized information center for coordination, each node independently implements the LKF to estimate its state. WSNs with stringent consensus requirements encounter challenges when utilizing MFDKF due to potential inconsistencies in the nodes' estimations. 
	
	To enhance consensus among nodes, a C-MFDKF is proposed by adding an average consensus term. The fusion rules for achieving consensus are defined as follows:
		\begin{equation} \label{consensus fusion}
			{\boldsymbol{\hat x}}_{n,k}^c = {{{\boldsymbol{\hat x}}}_{n,k}} + \eta \sum\nolimits_{m \in {A_n}} {({{{\boldsymbol{\hat x}}}_{m,k}} - {{{\boldsymbol{\hat x}}}_{n,k}}),}  
		\end{equation}
		where ${{\eta}}$ is the fusion gain of the node, ${\boldsymbol{\hat x}}_{n,k}^c$ is the final estimation of node $n$ after the consensus fusion. 
		The parameter $\eta$ determines the rate of consensus information propagation in the system. If the coefficient is too large, the system may become overly sensitive, leading to oscillations or excessive adjustments. On the other hand, if the coefficient is too small, the convergence speed may be too slow. According to Remark 3 and Theorem 2 from the literature \cite{olfati2007consensus}, the Perron matrix of a graph $\cal G$ with parameter $\eta$ should be a primitive matrix to ensure the convergence for the algorithm, finally the condition is:
		\begin{equation} \label{eta}
			0 < \eta  < 1/{({d_n})_{\max }}.
		\end{equation}
		
		Therefore, we propose the following parameter selection rules to ensure the convergence of the algorithm:
		\begin{equation}
			\eta  = \xi /{({d_n})_{\max }},
		\end{equation}
		where, $0 < \xi  < 1$. From Corollary 2 in Reference \cite{olfati2007consensus}, it follows that a larger parameter $\xi$ results in faster algorithm convergence, as long as the convergence condition $0 < \xi < 1$ is satisfied.
	
	Local coordination in the C-MFDKF employs the estimations of neighboring nodes to converge the estimations of all nodes in the WSN. The C-MFDKF is simply summarized in Algorithm \ref{C-MFDKF}. 
	\begin{algorithm}
		\caption{Consensus MFDKF}
		\label{C-MFDKF}
		Initialization: 
			$\chi _{m,0}^i = \alpha _m^i,{\boldsymbol{\hat x}}_{m,0}^i = {\boldsymbol{\hat x}}_{m,0}^c = {{{\boldsymbol{\hat x}}}_0},{\boldsymbol{\hat P}}_{m,0}^i = {{{\boldsymbol{\hat P}}}_0},\xi  = {\xi _0}$\;
		Choose the fusion gain ${\eta}$: (\ref{eta})\;
		\For{$k=1:T$}{\For{${m \in {{\cal A}_n}}$}{
				MFDKF steps: implement MFDKF as proposed in Algorithm \ref{MFDKF} form (\ref{fusion possibility}) to (\ref{state fusion})\;
			}
			Consensus fusion: update the estimations through the consensus fusion rules (\ref{consensus fusion});
		}
	\end{algorithm}
	
	The MFDKF algorithm requires independent state estimation for each sub-model, and its computational intensity increases with the number of sub-models. The number of sub-models grows rapidly with the number of nodes and noise components, as shown in equation (\ref{model number}). This adversely affects the real-time performance of both the MFDKF and C-MFDKF algorithms. Moreover, the node must exchange information about the observing system with their neighbors, leading to increased communication volume as the number of nodes grows. This poses challenges for the MFDKF and C-MFDKF algorithms in communication-constrained scenarios, such as underwater WSNs with limited bandwidth. Therefore, a simplified version of MFDKF is proposed incorporating the previously mentioned consensus fusion rules. The S-MFDKF algorithm requires fewer sub-models to be fused and reduces the number of interactions among nodes compared to the MFDKF and C-MFDKF algorithms. 
	
	In the S-MFDKF, each node implements the LKF solely using its own measurement, resulting in a communication-free LKF process. This approach differs from systems with robust communication capabilities, where nodes can also utilize measurements from neighboring sensors for filtering. Subsequently, the aforementioned consensus fusion rule is applied to the estimation of each node. This method aims to enhance the accuracy of each LKF estimation by fusing the estimations from the neighboring nodes. Algorithm \ref{S-MFDKF} provides a summary of the steps involved in the S-MFDKF. 
	\begin{algorithm}
		\caption{Simplified MFDKF}
		\label{S-MFDKF}
		Initialization: 
			$\chi _{m,0}^i = \alpha _m^i,{\boldsymbol{\hat x}}_{m,0}^i = {\boldsymbol{\hat x}}_{m,0}^c = {{{\boldsymbol{\hat x}}}_0},{\boldsymbol{\hat P}}_{m,0}^i = {{{\boldsymbol{\hat P}}}_0},\xi  = {\xi _0}$\;
		Choose the fusion gain ${\eta}$: (\ref{eta})\;
		\For{$k=1:T$}{\For{${m \in {{\cal A}_n}}$}{
				MFDKF steps: implement MFDKF as proposed in Algorithm \ref{MFDKF} form (\ref{fusion possibility}) to (\ref{state fusion}) in a communication-free way, i.e.
					\begin{equation}\label{L'}
						\begin{array}{l}
							{L_m} = {\kappa _m},{{{\boldsymbol{\tilde z}}}_{m,k}} = {{\boldsymbol{z}}_{m,k}},\\
							{{{\boldsymbol{\tilde H}}}_{m,k}} = {{\boldsymbol{H}}_{m,k}},{{{\boldsymbol{\tilde R}}}_{m,k}} = {{\boldsymbol{R}}_{m,k}};
						\end{array}
					\end{equation}}
			Consensus fusion: update the estimations through the consensus fusion rules (\ref{consensus fusion});}
	\end{algorithm}
	
	Table \ref{Communication contents} presents the communication contents and the number of sub-models about the proposed MFDKF and its derived algorithms. The S-MFDKF requires fewer sub-models to be fused and its computational effort is correspondingly lower. The detailed analysis is given in Section \ref{computational}. It can also be concluded that the S-MFDKF exhibits minimal information interaction with its neighbors, while the C-MFDKF demonstrates the highest level of interaction. The MFDKF, C-MFDKF, and S-MFDKF are equivalent when communication between nodes is prohibited.
	
	\begin{table}[!ht]
		\caption{Communication contents between the node and its neighbor in proposed algorithm}
		\centering
			\begin{tabular}{cccc}
				\hline
				Algorithm              & MFDKF & C-MFDKF & S-MFDKF \\ \hline
				Contents & $\begin{array}{l}
					{{\boldsymbol{H}}_{n,k}},
					{{\boldsymbol{z}}_{n,k}}\\
					{{\boldsymbol{R}}_{n,k}}
				\end{array}$     & $\begin{array}{l}
					{{\boldsymbol{H}}_{n,k}},
					{{\boldsymbol{z}}_{n,k}}\\
					{{\boldsymbol{R}}_{n,k}},
					{{{\boldsymbol{\hat x}}}_{n,k}}
				\end{array}$       & ${{{\boldsymbol{\hat x}}}_{n,k}}$       \\ \\ 
				Number & $\prod\limits_{{n_t} \in {{\cal A}_n}} {{\kappa _{{n_t}}}}$ & $\prod\limits_{{n_t} \in {{\cal A}_n}} {{\kappa _{{n_t}}}}$ & $\sum\limits_{{n_t} \in {{\cal A}_n}} {{\kappa _{{n_t}}}} $ \\
				\hline
		\end{tabular}
		\label{Communication contents}
	\end{table}
	
	\section{Performance Analysis} \label{performance}
	This section presents analyses to elucidate the performance of the proposed MFDKF and its derived algorithms, encompassing mean error, mean-square error, and computational complexity. Our framework is adaptable to time-varying settings while adhering to time-invariant assumptions to facilitate analysis:
	\begin{enumerate}
		\item The transition matrix and the measurement matrix of the system are time-invariant, thus ${{\boldsymbol{A}}_{n,k}} = {\boldsymbol{A}},{{\boldsymbol{H}}_{n,k}} = {\boldsymbol{H}}$.
		\item The state noise ${{\boldsymbol{w}}_k}$ and the measurement noise ${{\boldsymbol{v}}_{n,k}}$ are independent of each other, and the expectation of ${{\boldsymbol{w}}_k}$ is zero.  
	\end{enumerate}
	\subsection{Mean Error Analysis}
	The performance of mean error is reflected in the expectation of the estimation error $E\left( {{{\boldsymbol{\varepsilon }}_{n,k}}} \right)$. The following discussion is divided into two parts: the essential MFDKF algorithm and the consensus fusion rules utilized in C-MFDKF and S-MFDKF.  
	
	Firstly, the MFDKF algorithm is considered. The estimation error of node $n$ with the measurement model $j$ is determined by
	\begin{equation} \label{error_j}
		{\boldsymbol{\varepsilon }}_{n,k}^j = {{\boldsymbol{x}}_k} - {\boldsymbol{\hat x}}_{n,k}^j.
	\end{equation}
	According to (\ref{state fusion}), the LKF estimation error of node $n$ at time $k$ is 
	\begin{equation} \label{error_func}
			{{\boldsymbol{\varepsilon }}_{n,k}} = {{\boldsymbol{x}}_k} - {{{\boldsymbol{\hat x}}}_{n,k}} = \sum\nolimits_{j = 1}^{{L_n}} {\chi _{n,k}^j{\boldsymbol{\varepsilon }}_{n,k}^j} .
	\end{equation}
	Substituting (\ref{process_fun_1}), (\ref{fusion measurement}), (\ref{state_fusion_func}), (\ref{updating_funcj}) into (\ref{error_j}), we obtain
	\begin{equation} \label{error_ja}
		{\boldsymbol{\varepsilon }}_{n,k}^j \!\!=\!\! {\boldsymbol{f}}_{n,k}^j{\boldsymbol{A}}\!\!\sum\nolimits_{i = 1}^{{L_n}} \!\!{\pi _{n,k - 1}^{ij}{\boldsymbol{\varepsilon }}_{n,k - 1}^i}  \!+\! {\boldsymbol{f}}_{n,k}^j{{\boldsymbol{w}}_{k - 1}} \!-\! {\boldsymbol{K}}_{n,k}^j{\boldsymbol{\tilde v}}_{n,k}^j
	\end{equation}
	where
	\begin{equation} \label{error_jad}
		\begin{array}{*{20}{l}}
			{{\boldsymbol{f}}_{n,k}^j = {\boldsymbol{I}} - {\boldsymbol{K}}_{n,k}^j{\boldsymbol{\tilde H}}}\\
			{ = {\boldsymbol{I}} - {\boldsymbol{\bar P}}_{n,k}^j{{{\boldsymbol{\tilde H}}}^T}{{({\boldsymbol{\tilde H\bar P}}_{n,k}^j{{{\boldsymbol{\tilde H}}}^T} + {\boldsymbol{\tilde R}}_{n,k}^j)}^{ - 1}}{\boldsymbol{\tilde H}}}\\
			{ = {{\left[ {{{({\boldsymbol{\bar P}}_{n,k}^j)}^{ - 1}} + {{{\boldsymbol{\tilde H}}}^T}{{({\boldsymbol{\tilde R}}_{n,k}^j)}^{ - 1}}{\boldsymbol{\tilde H}}} \right]}^{ - 1}}{{({\boldsymbol{\bar P}}_{n,k}^j)}^{ - 1}}.}
		\end{array}
	\end{equation}
	
	The expectation of ${\boldsymbol{\varepsilon }}_{n,k}^j$ is then given by
	\begin{equation} \label{expoferr_j}
		E({\boldsymbol{\varepsilon }}_{n,k}^j) = {\boldsymbol{f}}_{n,k}^j{\boldsymbol{A}}\sum\nolimits_{i = 1}^{{L_n}} {\pi _{n,k - 1}^{ij}E({\boldsymbol{\varepsilon }}_{n,k - 1}^i)} .
	\end{equation}
	It is known that 
	%\begin{small}
	\begin{equation}
		\begin{array}{l}
			\left\| {\sum\nolimits_{i = 1}^{{L_n}} {\pi _{n,k - 1}^{ij}E({\boldsymbol{\varepsilon }}_{n,k - 1}^i)} } \right\| \le \\
			\max \left[ {\left\| {E({\boldsymbol{\varepsilon }}_{n,k - 1}^1)} \right\|\!,\!\left\| {E({\boldsymbol{\varepsilon }}_{n,k - 1}^2)} \right\|\!,\! \cdots\! ,\!\left\| {E({\boldsymbol{\varepsilon }}_{n,k - 1}^{{L_n}})} \right\|} \right]
		\end{array}
	\end{equation}
	%\end{small}
	and denote the right-hand side as $\left\| {E\left( {{{\bar \varepsilon }_{n,k - 1}}} \right)} \right\|$, the inequality is:
	\begin{equation}
		\left\| {\sum\nolimits_{i = 1}^{{L_n}} {\pi _{n,k - 1}^{ij}E({\boldsymbol{\varepsilon }}_{n,k - 1}^i)} } \right\| \le \left\| {E({{\bar \varepsilon }_{n,k - 1}})} \right\|.
	\end{equation}
	Substituting the inequality into (\ref{expoferr_j}), we have
	\begin{equation} \label{me_j}
		E\left( {{\boldsymbol{\varepsilon }}_{n,k}^j} \right) \le {\boldsymbol{f}}_{n,k}^j{\boldsymbol{A}}\left\| {E\left( {{{{\boldsymbol{\bar \varepsilon }}}_{n,k - 1}}} \right)} \right\|.
	\end{equation}
	
	The matrices ${{\boldsymbol{\bar P}}_{n,k}^{j}}$ and ${{{{\boldsymbol{\tilde H}}}^T}{{\left( {{\boldsymbol{\tilde R}}_{n,k}^j} \right)}^{ - 1}}{\boldsymbol{\tilde H}}}$ being positive semi-definite,  ensure the stability of ${\boldsymbol{f}}_{n,k}^{j}$. Moreover, the stability of ${\boldsymbol{A}}$ in (\ref{me_j}) guarantees the stability of the estimation of model $j$, ${{\boldsymbol{\varepsilon }}_{n,k}^j}$. From (\ref{error_func}), the expectation of estimation error is
	\begin{equation}
		E({{\boldsymbol{\varepsilon }}_{n,k}}) = \sum\nolimits_{j = 1}^{{L_n}} {\chi _{n,k}^jE({\boldsymbol{\varepsilon }}_{n,k}^j)} .
	\end{equation}
	This equation signifies the stability of the estimation error for node $n$, ${{{\boldsymbol{\varepsilon }}_{n,k}}}$ is stable.
	
	Then, the performance of consensus fusion rules is considered. Following the consensus fusion rules, the expectation of estimation error is 
	\begin{equation} \label{error of consensus}
		E\left( {{\boldsymbol{\varepsilon }}_{n,k}^c} \right) = E\left( {{{\boldsymbol{x}}_k} - {\boldsymbol{\hat x}}_{n,k}^c} \right).
	\end{equation}
	By substituting (\ref{consensus fusion}) (\ref{error_func}) into (\ref{error of consensus}) we obtain
	\begin{equation} \label{error2 fusion}
		E\left( {{\boldsymbol{\varepsilon }}_{n,k}^c} \right) = E\left[ {{{\boldsymbol{\varepsilon }}_{n,k}} + \eta {\boldsymbol{A}}\sum\nolimits_{m \in {A_n}}  \left( {{{\boldsymbol{\varepsilon }}_{m,k}} - {{\boldsymbol{\varepsilon }}_{n,k}}} \right)} \right].
	\end{equation}
	
	As ${{\eta _n}}$ is a constant, ${\boldsymbol{A}}$ is time-invariant, and the stability of ${{{\boldsymbol{\varepsilon }}_{n,k}}}$ has been previously established, we can conclude that ${{\boldsymbol{\varepsilon }}_{n,k}^c}$ is also stable.
	
	\subsection{Mean-Square Error Analysis}
	Firstly, the basic MFDKF is analyzed. The performance of mean-square error is determined by the estimation error covariance matrix, which is given by
	\begin{equation} \label{error_covfun}
		E({{\boldsymbol{\varepsilon }}_{n,k}}{\boldsymbol{\varepsilon }}_{n,k}^T) \!=\!\! \sum\nolimits_{i = 1}^{{L_n}}\! {\sum\nolimits_{j = 1}^{{L_n}} {\chi _{n,k}^i/\chi _{n,k}^jE[{\boldsymbol{\varepsilon }}_{n,k}^i{{({\boldsymbol{\varepsilon }}_{n,k}^j)}^T}]} } .
	\end{equation}
	Assuming that,
	\begin{equation}
		\begin{array}{*{20}{l}}
			{E[{\boldsymbol{\varepsilon }}_{n,k}^i{{({\boldsymbol{\varepsilon }}_{n,k}^j)}^T}] = 0,i \ne j,}\\
			{{\boldsymbol{\Omega }}_{n,k}^i = E[{\boldsymbol{\varepsilon }}_{n,k}^i{{({\boldsymbol{\varepsilon }}_{n,k}^j)}^T}],{{\boldsymbol{\Omega }}_{n,k}} = E({{\boldsymbol{\varepsilon }}_{n,k}}{\boldsymbol{\varepsilon }}_{n,k}^T).}
		\end{array}
	\end{equation}
	Equation (\ref{error_covfun}) can be expressed as
	\begin{equation} \label{error2}
		{{\boldsymbol{\Omega }}_{n,k}} = \sum\nolimits_{j = 1}^{{L_n}} {{\boldsymbol{\Omega }}_{n,k}^j} .
	\end{equation}
	From the representation in (\ref{error_ja}) (\ref{error_jad}), ${\boldsymbol{\Omega }}_n^j\left( k \right)$ is calculated by
	\begin{equation} \label{omaga_funcj}
		{\boldsymbol{\Omega }}_{n,k}^j = {\boldsymbol{\Gamma }}_{n,k}^j\sum\nolimits_{j = 1}^{{L_n}} {{\boldsymbol{\Omega }}_{n,k - 1}^i} {({\boldsymbol{\Gamma }}_{n,k}^j)^T} + {\boldsymbol{\Theta }}_{n,k}^j,
	\end{equation}
	where, 
	\begin{equation}
		\begin{array}{l}
			{\boldsymbol{\Theta }}_{n,k}^j = {\boldsymbol{f}}_{n,k}^j{{\boldsymbol{Q}}_k}{({\boldsymbol{f}}_{n,k}^j)^T} + {\boldsymbol{K}}_{n,k}^j{\boldsymbol{\tilde R}}_{n,k}^j{({\boldsymbol{K}}_{n,k}^j)^T},\\
			{\boldsymbol{\Gamma }}_{n,k}^j = {\boldsymbol{f}}_{n,k}^j{\boldsymbol{A}}
		\end{array}
	\end{equation}
	
	Assuming that ${\boldsymbol{Q}}_k$, and ${{\boldsymbol{\tilde R}}_{n,k}^j}$ are stationary \cite{talebi2016distributed}, ${{\boldsymbol{f}}_{n,k}^j}$, ${{\boldsymbol{\Theta }}_{n,k}^j}$, and ${{\boldsymbol{\Gamma }}_{n,k}^j}$ are time-invariant. As a result, ${\boldsymbol{\Omega }}_{n,k}^j$  converges. The time-invariant variables can be summarized as
	\begin{equation}
		\begin{array}{*{20}{l}}
			{\mathop {\lim }\limits_{k \to \infty } {\boldsymbol{\Omega }}_{n,k}^j \!=\! {\boldsymbol{\Omega }}_n^j,}
			{\mathop {\lim }\limits_{k \to \infty } {\boldsymbol{\Gamma }}_{n,k}^j \!=\! {\boldsymbol{\Gamma }}_n^j,}
			{\mathop {\lim }\limits_{k \to \infty } {\boldsymbol{\Theta }}_{n,k}^j \!=\! {\boldsymbol{\Theta }}_n^j.}
		\end{array}
	\end{equation}
	Taking the limit of (\ref{omaga_funcj}), 
	\begin{equation}\label{omaga_limit}
		{\boldsymbol{\Omega }}_n^j = {L_n}{\boldsymbol{\Gamma }}_n^j{\boldsymbol{\Omega }}_n^j{\left( {{\boldsymbol{\Gamma }}_n^j} \right)^T} + {\boldsymbol{\Theta }}_n^j.
	\end{equation}
	Applying the rules of matrix vectorization as follows
	\begin{equation}
		\begin{array}{*{20}{l}}
			{vec\left( {{{\boldsymbol{\varsigma }}_1}{{\boldsymbol{\varsigma }}_2}{{\boldsymbol{\varsigma }}_3}} \right) = \left( {{\boldsymbol{\varsigma }}_3^T \otimes {{\boldsymbol{\varsigma }}_1}} \right)vec\left( {{{\boldsymbol{\varsigma }}_2}} \right),}\\
			{vec\left( {{{\boldsymbol{\varsigma }}_1} + {{\boldsymbol{\varsigma }}_2}} \right) = vec\left( {{{\boldsymbol{\varsigma }}_1}} \right) + vec\left( {{{\boldsymbol{\varsigma }}_2}} \right),}
		\end{array}
	\end{equation} 
	(\ref{omaga_limit}) can be rewritten as 
	\begin{equation} \label{omaga_vector}
		vec\left( {{\boldsymbol{\Omega }}_n^j} \right) = {L_n}{\boldsymbol{\Gamma }}_n^j \otimes {\boldsymbol{\Gamma }}_n^jvec\left( {{\boldsymbol{\Omega }}_n^j} \right) + vec\left( {{\boldsymbol{\Theta }}_n^j} \right).
	\end{equation}
	The closed-form solution for (\ref{omaga_vector}) can be obtained as follows:
	\begin{equation}
		vec\left( {{\boldsymbol{\Omega }}_n^j} \right) = {\left( {{\boldsymbol{I}} - {L_n}{\boldsymbol{\Gamma }}_n^j \otimes {\boldsymbol{\Gamma }}_n^j} \right)^{ - 1}}vec\left( {{\boldsymbol{\Theta }}_n^j} \right).
	\end{equation}
	
	Equation (\ref{error2}) reveals that the estimation error covariance matrix ${{\boldsymbol{\Omega }}_{n,k}}$ is obtained by summing the estimation error covariance matrix of sub-models ${{\boldsymbol{\Omega }}_{n,k}^j}$. Thus, a closed solution exists for the estimation error covariance matrix ${{\boldsymbol{\Omega }}_{n,k}}$.
	
	The mean-square error of the consensus fusion rule is then considered. According to (\ref{error2 fusion}) the estimation error covariance matrix is written as
	\begin{equation}
		E[{\boldsymbol{\varepsilon }}_{n,k}^c{({\boldsymbol{\varepsilon }}_{n,k}^c)^T}\!] = {{\boldsymbol{\Omega }}_{n,k}} + {\boldsymbol{A}}\sum\nolimits_{m \in {A_n}}\!\!\!\!{[{{\boldsymbol{\Omega }}_{m,k}} + {{\boldsymbol{\Omega }}_{n,k}}]} {{\boldsymbol{A}}^T}.
	\end{equation}
	Its closed-form solution depends on ${{\boldsymbol{\Omega }}_{n,k}}$, whose closed-form solution exists.
	
	\subsection{Computational Complexity} \label{computational}
	This section analyzes the computational complexity of the proposed algorithms in terms of floating-point operations, following the methodology described in \cite{chen2019minimum}. Table \ref{Computational} presents the computational complexities of the equations, with $d$ and $L$ representing the number of neighbors and measurement sub-models about the LKF, respectively. The computational complexity of CDKF relies on equations (\ref{CDKF_prediction})-(\ref{CDKF Mk}), while the computational complexity of MFDKF depends on (\ref{fusion possibility})-(\ref{state fusion}). Accordingly, the computational complexity of CDKF and MFDKF are respectively,
	\begin{equation}
		{S_{CDKF}} = 6{p^3} + 6d{p^2}q + 4{d^2}p{q^2} + dpq - p + O( {{d^3}{q^3}} ),
	\end{equation}
		\begin{equation} \label{jsl MFDKF}
			{S_{MFDKF}} \!=\! \left[\!\! \begin{array}{l}
				{S_{CDKF}} + 2{d^2}{q^2} + \left( {4L \!-\! 1} \right){p^2} + \\
				\left( {3L \!-\! 1} \right)p + dq + q + 4L + d + 4
			\end{array} \!\! \right]L \!-\! p.
	\end{equation}
	It can be obtained that ${S_{MFDKF}} > {S_{CDKF}}$. Meanwhile, an increase in $L$ leads to a sharp increase in the computation of MFDKF.
\begin{table}[!ht]
	\caption{Computational complexities of equations}
	\centering
	\begin{tabular}{lll}
		\hline
		Equation & $ + $ and $ \times $ & \begin{tabular}[c]{@{}l@{}}Division, matrix inverse, \\ Cholesky decomposition \\ and exponentiation\end{tabular} \\  \hline
		(\ref{CDKF_prediction})        & $2{p^2} - p$                                              & $0$                                                                                                                                     \\
		(\ref{CDKF_P_})        & $4{p^3} - {p^2}$                                              & $0$                                                                                                                                     \\
		(\ref{S_c})        & $2d{p^2}q + 2{d^2}p{q^2} - dpq$                                              & $O\left( {{d^3}{q^3}} \right)$                                                                                                                                     \\
		(\ref{K_c})        & $2d{p^2}q + 2{d^2}p{q^2} - 2dpq$                                              & $0$                                                                                                                                     \\
		(\ref{v_c})       & $2dpq$                                              & $0$                                                                                                                                     \\
		(\ref{xk})       & $2dpq$                                              & $0$                                                                                                                                     \\
		(\ref{CDKF Mk})       & $2{p^3} + 2d{p^2}q - {p^2}$                                              & $0$                                                                                                                                     \\
		(\ref{fusion possibility})       & $3{L^2} - L$                                              & $0$                                                                                                                                     \\
		(\ref{state_fusion_func})       & $2{L^2}p - Lp$                                              & $0$                                                                                                                                     \\
		(\ref{cov_xpre})       & $4{L^2}{p^2} - L{p^2} + {L^2}p$                                              & $0$                                                                                                                                     \\
		(\ref{updating_funcj})       & $\begin{array}{l}
			(4{p^3} + {p^2} + 4d{p^2}q + \\
			4{d^2}p{q^2} + dpq - p + 1)L
		\end{array}$                                              & $LO\left( {{d^3}{q^3}} \right)$                                                                                                                                     \\
		(\ref{p_hat_j})       & $\left( {2{p^3} + 2d{p^2}q - {p^2}} \right)L$                                              & $0$                                                                                                                                     \\
		(\ref{likelihood})       & $\left( {2{d^2}{q^2} + dq + 1} \right)L$                                              & $4L + LO\left( {{d^3}{q^3}} \right)$                                                                                                                                                                                                                                                                          \\
		(\ref{proba_update})       & ${L^2} - L$                                              & $L$                                                                                                                                     \\
		(\ref{state fusion})       & $2Lp - p$                                              & $0$                                                                                                                                     \\ 
		(\ref{consensus fusion}) & $2p^2d + 2p$ & $0$ \\ \hline
	\end{tabular}
	\label{Computational}
\end{table}

	Compared to MFDKF, C-MFDKF requires the additional computation of (\ref{consensus fusion}), and its computational complexity can be expressed as
		\begin{equation} \label{computation_cmfdkf}
			{S_{C - MFDKF}} = {S_{MFDKF}} + 2{p^2}d + 2p.
		\end{equation}
		One can conclude that ${S_{C - MFDKF}} > {S_{MFDKF}}$.
	
	The process of S-MFDKF can be viewed as applying consensus fusion after MFDKF without communication and its computational complexity can be expressed as
	\begin{equation} \label{computation_smfdkf}
		{S_{S - MFDKF}} = {S_{MFDKF|d = 1,L = L'}} + 2{p^2}d + 2p + 1.
	\end{equation}
	Here, $L'$ is the number of submodels to be fused in S-MFDKF. According to (\ref{model number}) and (\ref{L'}), $L' \le L$. The number of neighbors $d \ge 1$, thus ${S_{S - MFDKF}} \le {S_{MFDKF}}$.
	
	From \cite{hu2022efficient}, we can obtain the computational complexity of DMCKF, denoted as
	\begin{equation}
		\begin{array}{l}
			{S_{DMCKF}} \!=\! \left( {2\tau \!+\! 8} \right){p^3} \!+\! \left( {4\tau \!+\! 6} \right){p^2}dq + \left( {2\tau \!-\! 1} \right){p^2}\\
			+ \left( {4\tau + 2} \right)p{d^2}{q^2} + \left( {3\tau - 1} \right)pdq + \left( {4\tau - 1} \right)p\\
			+ 2{\tau}{d^3}{q^3} + 2{\tau}dq + {\tau}O\left( {{p^3}} \right) + 2{\tau}O\left( {{d^3}{q^3}} \right).
		\end{array}
	\end{equation}
	The computational complexity of DMEEKF, as stated in \cite{feng2023distributed}, is given by
	\begin{equation}
		\begin{array}{l}
			{S_{DMEEKF}} = \left( {8{\tau} + 8} \right){p^3} + 8{\tau}{q^3} + \left( {22{\tau} + 6} \right){p^2}q\\
			- \left( {2{\tau} + 1} \right){p^2} + \left( {18{\tau} + 2} \right)p{q^2} + {\tau}q + \left( {5{\tau} - 1} \right)p\\
			+ \left( {10{\tau} - 1} \right)pq + {\tau}O\left( {{p^3}} \right) + {\tau}O\left( {{q^3}} \right) + {\tau}O\left( {pq} \right).
		\end{array}
	\end{equation}
	Here, $\tau$ represents the average fixed-point iteration number of node $n$. Its value is related to the value taken for the kernel width, the smaller the kernel width the larger the $\tau$. The average simulation time of the above algorithms is shown in Section \ref{simulations}, which reflects that the computational complexity of MFDKF is larger than that of DMCKF and DMEEKF. 
	\section{Simulations} \label{simulations}
	This section presents simulations to validate the performance of the proposed MFDKF and its derived algorithms. The simulations are executed considering a linear system. 
	In Section \ref{simulation 1}, the parameter is first set as its minimum value $\kappa_n=2$, while compared with following algorithms:
		\begin{enumerate}
			\item CDKF \cite{he2020distributed}: A DKF that can achieve minimum mean square error estimation in the presence of Gaussian noise.
			\item DMCKF \cite{wang2019distributed}: The ITL-based DKF outperforms the CDKF under the non-Gaussian noise as it can capture higher-order moment information of the error and adjust the estimation accordingly.
			\item DMEEKF \cite{feng2023distributed}: A DKF based on the MEE criterion that performs better than DMCKF under non-Gaussian noise.
			\item BIKF \cite{fan2022background}: A KF effectively addresses impulsive noise by utilizing two components to represent the predominant noise and the occasional impulsive noise, respectively.
	\end{enumerate}
	It is worth noting that the performance of the algorithms DMCKF and DMEEKF is influenced by the parameter kernel width $\sigma$. Therefore, we have selected the most appropriate $\sigma$ for comparison to ensure a fair evaluation of these algorithms. 
	
	The value of parameter $\kappa_n$ is then discussed. Section \ref{simulation 2} focuses on specific scenarios, where the performance of the derived C-MFDKF and S-MFDKF is tested. 
	
	Unless otherwise specified, each simulation in this paper consists of $500$ independent Monte Carlo runs, and the root mean square error (RMSE) of the estimations is calculated using $1000$ samples. The RMSE to measure the accuracy is calculated by 
	\begin{equation}
		RMSE(m) = \sqrt {1/M\sum\nolimits_{k = 1}^T {{{\left\| {{{{\boldsymbol{\hat x}}}_k} - {{\boldsymbol{x}}_k}} \right\|}^2}} } ,
	\end{equation}
	where ${{{{\boldsymbol{\hat x}}}_k}}$ and ${{{\boldsymbol{x}}_k}}$ respectively denote the estimated state and true state in $k$-th run. $M$ is the number of Monte Carlo runs, and $T$ is the number of samples. To highlight the performance differences between algorithms, we use decibels as the unit to represent the RMSE in the figures \cite{ZHONG2024111343}:
	\begin{equation}
			RMSE(dB) = 20\log \left( {{{RMSE\left( m \right)} \mathord{\left/
						{\vphantom {{RMSE\left( m \right)} {1\left( m \right)}}} \right.
						\kern-\nulldelimiterspace} {1\left( m \right)}}} \right).
	\end{equation}
	
	Simulations are implemented under Gaussian noise and some non-Gaussian noises as shown below.
	\begin{enumerate}
		\item The noise obeying Gaussian distribution is described as
		\begin{equation}
			{\boldsymbol{v}} \sim {\cal N}\left( {{\boldsymbol{\mu }},{\boldsymbol{\sigma }}} \right),
		\end{equation}
		where ${\cal N}\left( {{\boldsymbol{\mu }},{\boldsymbol{\sigma }}} \right)$ represents Gaussian distribution with mean ${\boldsymbol{\mu }}$ and variance ${\boldsymbol{\sigma }}$.
		\item  The noise obeying mixed Gaussian distribution is described as
			\begin{equation} 
				{\boldsymbol{v}}\sim\lambda N\left( {{{\boldsymbol{\mu }}_1},{{\boldsymbol{\sigma }}_1}} \right) + \left( {1 - \lambda } \right)N\left( {{{\boldsymbol{\mu }}_2},{{\boldsymbol{\sigma }}_2}} \right),
		\end{equation}
		where $\lambda$ is the mixed coefficient of two Gaussian noises. The noise obeys this kind of mixed Gaussian distribution is denoted as ${\boldsymbol{v}}\sim M\left( {\lambda ,{{\boldsymbol{\mu }}_1},{{\boldsymbol{\mu }}_2},{{\boldsymbol{\sigma }}_1},{{\boldsymbol{\sigma }}_2}} \right)$.
		\item The $\alpha$-stable distribution is described as the following characteristic function
		\begin{equation}
			\phi \left( t \right) \!=\! \exp \left\{ {j\varpi t \!-\! \zeta {{\left| t \right|}^a}\left[ {1 \!+\! jb{\mathop{\rm sign}\nolimits} \left( t \right)\omega \left( {t,a} \right)} \right]} \right\},
		\end{equation}
		with
		\begin{equation}
			\omega \left( {t,a} \right) = \left\{ {\begin{array}{*{20}{l}}
					{\tan \left( {\frac{{a\pi }}{2}} \right),a \ne 1,}\\
					{\frac{2}{\pi }\log \left| t \right|,a = 1.}
			\end{array}} \right.
		\end{equation}
		$a$ denotes the characteristic exponent, $b$ is the symmetry parameter, $\zeta $ and $\varpi$ severally represent the dispersion parameter and the location parameter. $\left|  \cdot  \right|$ in this function denotes the absolute value operation. The noise that follows this kind of distribution is expressed as ${\boldsymbol{v}} \sim {\cal S}\left( {a,b,\zeta,\varpi } \right)$.
	\end{enumerate}
	
\begin{figure}[!ht]
		\centering
		\includegraphics[width=0.9\linewidth]{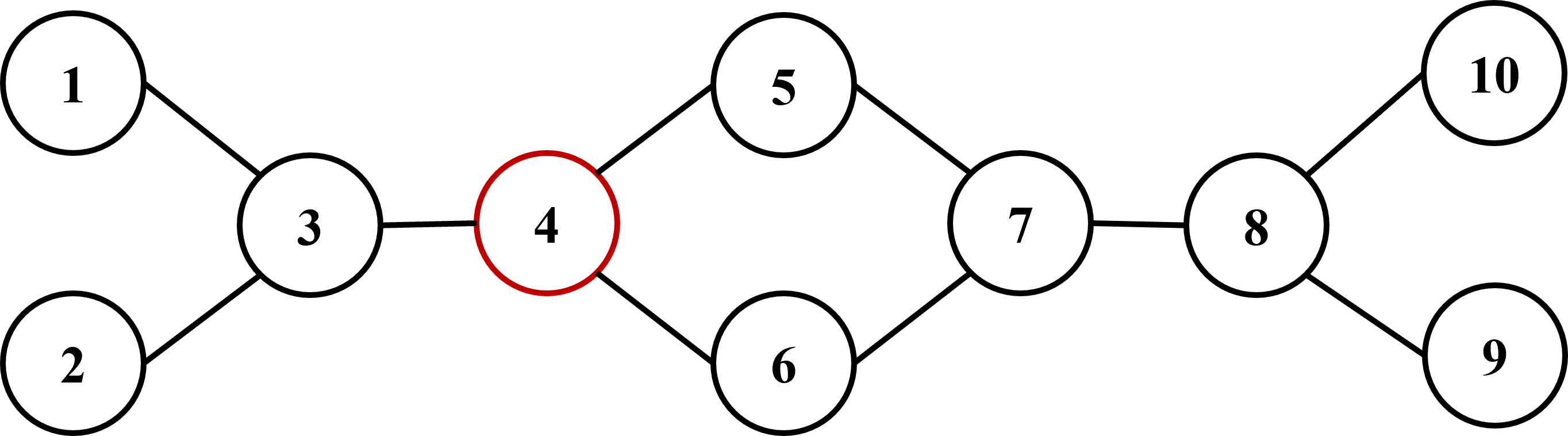}
		\caption{Communication typology of the WSN}
		\label{mao3}
	\end{figure}

	The simulations are conducted on a WSN with a communication topology illustrated in Fig. \ref{mao3}. Node $4$, which has nodes $3$, $5$, and $6$ as neighbors, is selected to compare the performance of the algorithms.
	\subsection{Simulation of MFDKF} \label{simulation 1}
		In this section, the following target tracking system \cite{CHEN2019228} is considered:
		\begin{equation}
			\begin{array}{l}
				{{\boldsymbol{x}}_k} = {{\boldsymbol{A}}_{k - 1}}{{\boldsymbol{x}}_{k - 1}} + {{\boldsymbol{G}}_{k - 1}}{{\boldsymbol{w}}_{k - 1}},\\
				{{\boldsymbol{z}}_{n,k}} = {{\boldsymbol{H}}_{n,k}}{{\boldsymbol{x}}_k} + {{\boldsymbol{v}}_{n,k}},
			\end{array}
		\end{equation}
		where 
		\begin{equation}
			\begin{array}{*{20}{l}}
				{{{\boldsymbol{A}}_k} = \left[ {\begin{array}{*{20}{c}}
							1&{{s_k}}&0&0\\
							0&1&0&0\\
							0&0&1&{{s_k}}\\
							0&0&0&1
					\end{array}} \right],{{\boldsymbol{G}}_k} = \left[ {\begin{array}{*{20}{c}}
							{s_k^2/2}\\
							{{s_k}}\\
							{s_k^2/2}\\
							{{s_k}}
					\end{array}} \right],}\\
				{{{\boldsymbol{H}}_{n,k}} = \left[ {\begin{array}{*{20}{c}}
							1&0&0&0\\
							0&0&1&0
					\end{array}} \right],{s_k} = 0.3 + 0.2\sin (k),}
			\end{array}
		\end{equation}
		and $s_k$ is the time-varying sampling period, ${{\boldsymbol{w}}_k} \sim {\cal N}\left( {0,0.1} \right)$.
	
	The number of Gaussian components $\kappa_n = 2$ is considered. We set the measurement noise of each node as following $\alpha$-stable noise 
		\begin{equation}\label{alpha noise}
			{{\boldsymbol{v}}_{n,k}}\sim{\cal S}\left( {1.2,0,2,0} \right).
		\end{equation}
		\begin{figure}[!ht]
			\centering
			\includegraphics[width=0.9\linewidth]{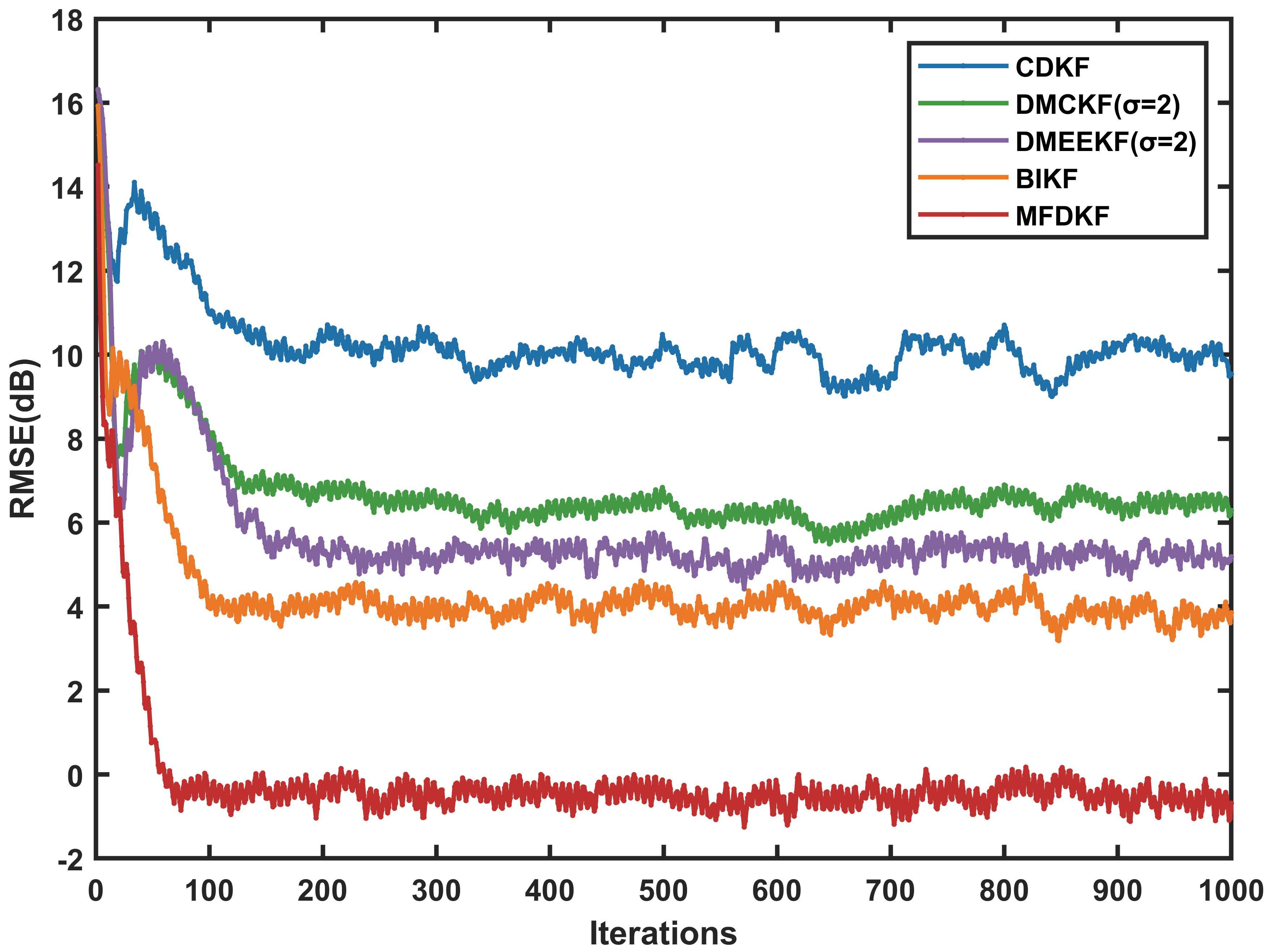} 
			\caption{Comparison of RMSEs under $\alpha$-stable noises, as represented by (\ref{alpha noise}). The kernel widths of DMCKF and DMEEKF are set to $2$.}
			\label{mao4}
	\end{figure}Fig. \ref{mao4} illustrates the RMSE performance of the proposed algorithm. It can be concluded that the MFDKF achieves the estimation with the lowest RMSEs. In descending order of their RMSE performance, the remaining algorithms are BIKF, DMEEKF, DMCKF, and CDKF.
	\begin{figure}[!ht]
		\centering
		\includegraphics[width=0.9\linewidth]{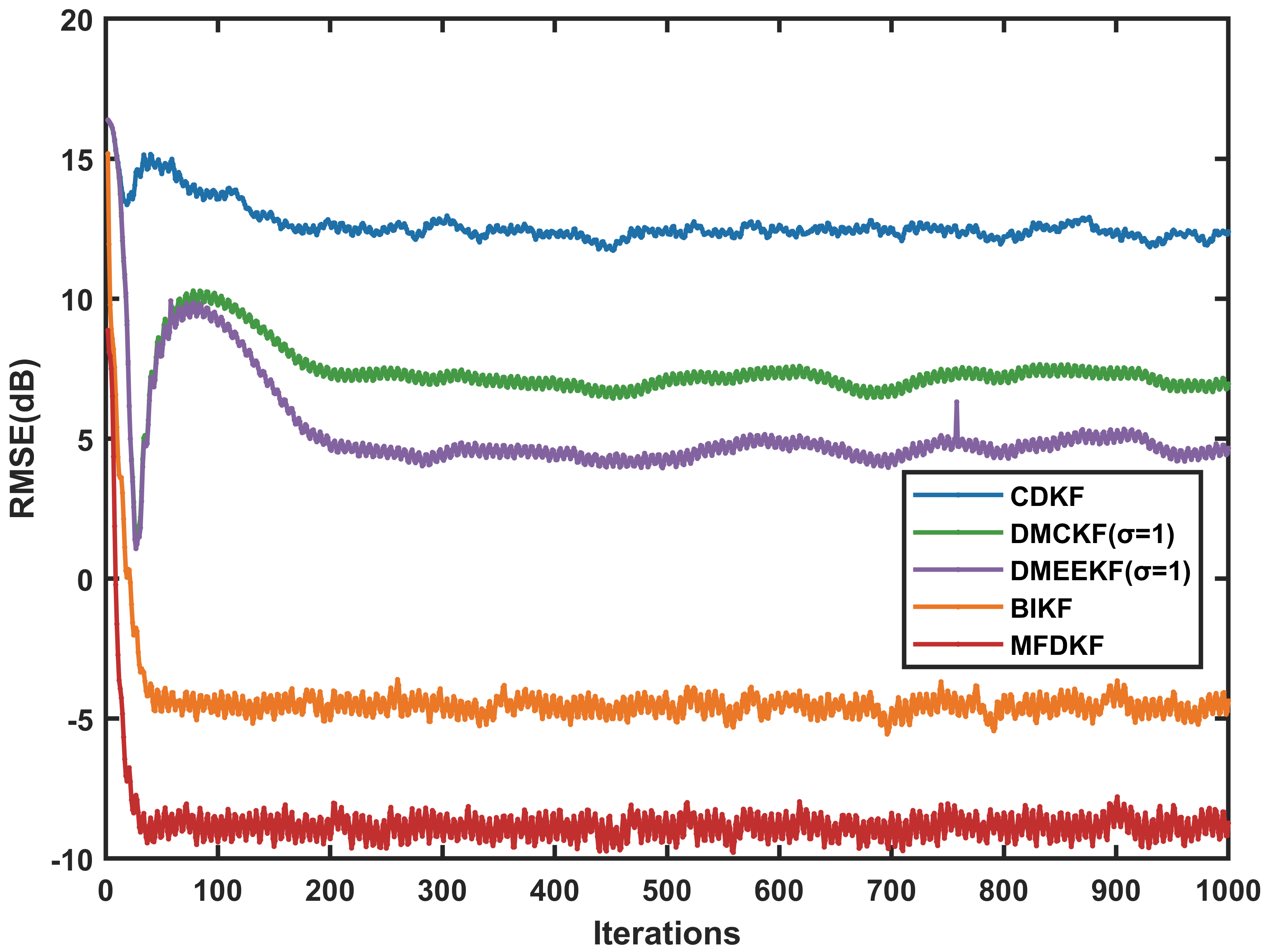}
		\caption{Comparison of RMSEs under mixed Gaussian noises, as represented by (\ref{eq:mixed Gaussian noise}). The kernel widths of DMCKF and DMEEKF are set to $1$.}
		\label{mao5}
	\end{figure}
	
	Subsequently, the measurement noise is set as following mixed Gaussian distribution
		\begin{equation} \label{eq:mixed Gaussian noise}
			{{\boldsymbol{v}}_{n,k}}\sim{\cal M}\left( {0.9,0,0,{{100}^2},1} \right).
		\end{equation}
		Fig. \ref{mao5} shows the MFDKF works stably with the best accuracy.
	
	\begin{figure}[!ht]
		\centering
		\includegraphics[width=0.9\linewidth]{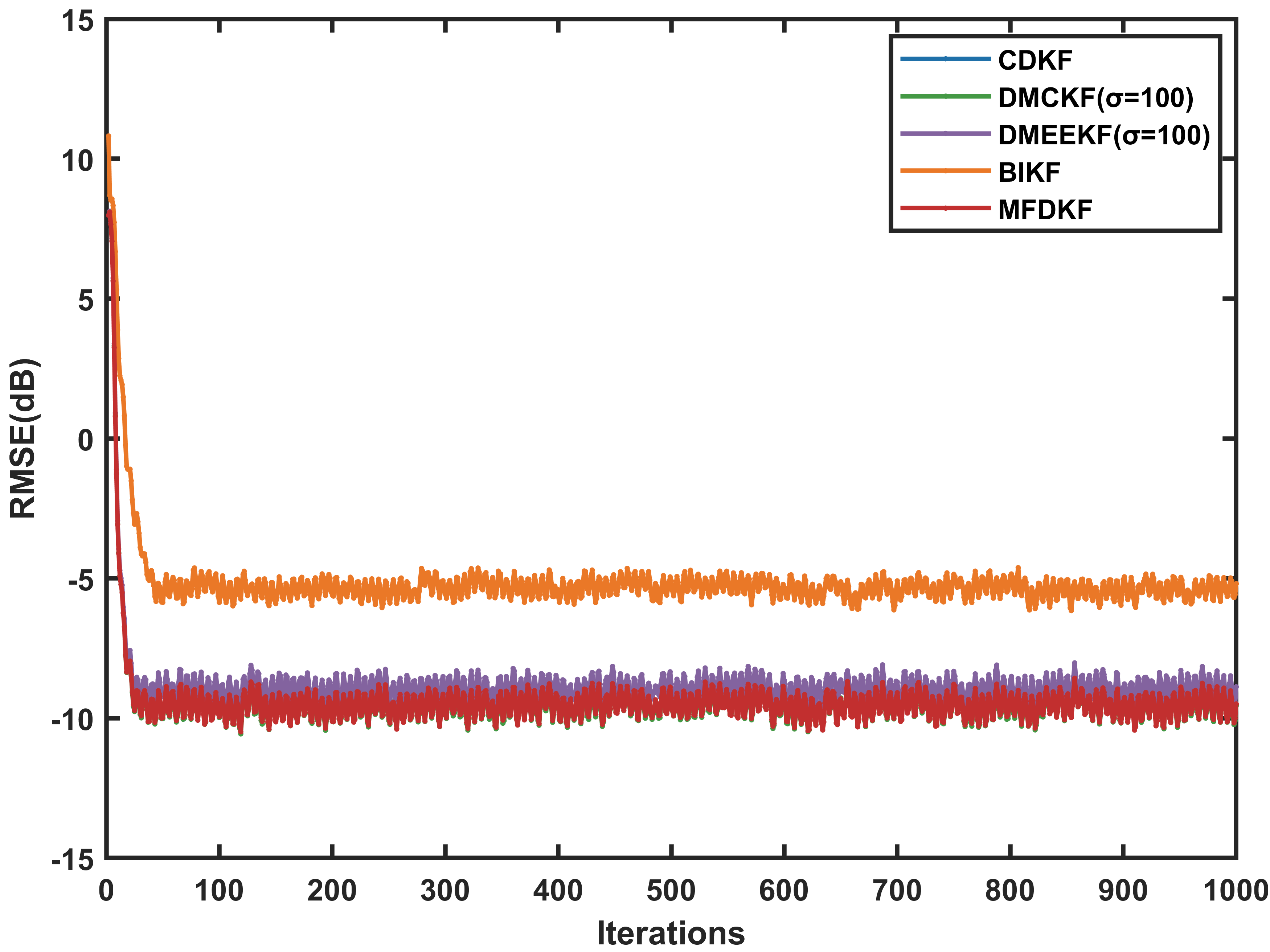} 
		\caption{Comparison of RMSEs under Gaussian noises, as represented by (\ref{Gaussian noise}). The kernel widths of DMCKF and DMEEKF are set to $100$.}
		\label{mao6}
	\end{figure}
	The following Gaussian noise is also considered
	\begin{equation} \label{Gaussian noise}
		{{\boldsymbol{v}}_{n,k}}\sim{\cal N}\left( {0,1} \right).
	\end{equation}
	As shown in Fig. \ref{mao6}, the performance of MFDKF is as good as CDKF, DMCKF, and DMEEKF, while BIKF performs less outstandingly. 
	\begin{table*}[!ht]
		\caption{The simulation time under different non-Gaussian noise}
		\centering
			\begin{tabular}{cccccccc} 
				\hline
				\multirow{2}*{Noises distribution} & \multicolumn{7}{c}{Average simulation time of the algorithms (s)}  \\
				& BIKF & CDKF & DMCKF & DMEEKF & MFDKF & C-MFDKF & S-MFDKF  \\ 
				\hline
				${\cal N}\left( {0,1} \right)$ & 0.0233 & 0.0181 &0.0361  &0.1035  & 0.5726  & 0.5926  & 0.0647  \\
				${\cal S}\left( {1.2,0,2,0} \right)$ &0.0278 &0.0192 &0.037  &0.1272  &0.5686  & 0.5967  & 0.0647  \\
				${\cal M}\left( {0.9,0,100^2,1} \right)$  & 0.0214   & 0.0191 &0.0379    & 0.1235    & 0.5746  & 0.5424  & 0.0656       \\
				\hline
		\end{tabular}
		\label{Simulation time}
	\end{table*}
	\begin{table*}[!ht]
		\caption{The RMSEs under different non-Gaussian noise}
		\centering
			\begin{tabular}{cccccccc} 
				\hline
				\multirow{2}*{Noises distribution} & \multicolumn{7}{c}{RMSE of the algorithms (m)}  \\
				& CDKF & DMCKF & DMEEKF & BIKF & MFDKF & C-MFDKF  & S-MFDKF    \\ 
				\hline
				${\cal N}\left( {0,1} \right)$ & 0.35 & 0.35 ($\sigma=100)$ & 0.37 ($\sigma=100$) & 0.57 & 0.35 & 0.33 & 0.49 \\
				${\cal S}\left( {1.2,0,2,0} \right)$ & 3.29 & 2.2 ($\sigma=2$) & 1.97 ($\sigma=2$) & 1.67 & 1 & 0.93 & 1.48 \\
				${\cal S}\left( {1.2,1,2,1} \right)$    &3.73   & 3.52($\sigma=9$)    & 3.63 ($\sigma=10$)    & 5.28    & 1 & 0.93 & 1.49     \\
				${\cal S}\left( {1.2,0.5,2,1} \right)$  & 3.35   & 2.35 ($\sigma=6$)    & 2.27 ($\sigma=3$)    & 2.46   & 1 & 0.91 & 1.44    \\
				${\cal S}\left( {1.5,0.5,1,1} \right)$ & 1.64   & 1.24 ($\sigma=1$)    & 1.26 ($\sigma=1$)    & 1.36    & 0.51 & 0.49 & 0.73       \\
				${\cal S}\left( {1.5,0.5,2,1} \right)$ & 1.99   &1.35 ($\sigma=1$)    & 1.51 ($\sigma=1$)    & 1.58    & 0.84 & 0.81  & 1.2      \\
				${\cal M}\left( {0.9,0,0,100^2,1} \right)$ & 4.3   & 2.39 ($\sigma=1$)    & 1.88 ($\sigma=1$)    & 0.62    & 0.38 & 0.35 & 0.52        \\
				${\cal M}\left( {0.9,0.5,1,100^2,1} \right)$  & 4.53   & 2.81 ($\sigma=1$)    & 2.38 ($\sigma=1$)    & 1.56    & 0.38 & 0.35 & 0.54     \\
				${\cal M}\left( {0.9,0.5,1,50^2,1} \right)$ & 2.59   & 1.47 ($\sigma=1$)    & 1.18 ($\sigma=1$)    & 0.63    & 0.38 & 0.36 & 0.54       \\
				${\cal M}\left( {0.8,0,1,50^2,5^2} \right)$ & 3.38   & 2.33 ($\sigma=1$)    & 2.3 ($\sigma=1$)    & 2.3    & 1.36 & 1.28 & 1.86       \\
				${\cal M}\left( {0.8,0,0.5,100^2,5^2} \right)$ & 5.5  & 3.42 ($\sigma=1$)    & 3.05 ($\sigma=1$)     & 2.28    & 1.35  & 1.26 & 1.94     \\ \hline
		\end{tabular}
		\label{tab:4}
	\end{table*}
	
	Average simulation times are listed in Table \ref{Simulation time} to further investigate the algorithm's computational complexity. Simulations were implemented in MATLAB R2020a and executed on an i5-12490F CPU with a clock speed of 3.0 GHz. It can be concluded that although MFDKF performs better, it is computationally burdensome. 
	
	To demonstrate the effectiveness of the proposed MFDKF, we evaluate its performance under various non-Gaussian measurement noise, including $\alpha$-stable and mixed Gaussian noise. The specific results of the simulations mentioned above are presented in Table \ref{tab:4}. It is evident from the results that the MFDKF surpasses other algorithms in the given non-Gaussian measurement scenarios.
	
	\begin{table}[!ht]
		\caption{RMSEs of different nodes in the WSN}
		\resizebox{\linewidth}{!}{
			\begin{tabular}{ccccccccccc}
				\hline
				Node & 1    & 2    & 3    & 4    & 5    & 6    & 7    & 8    & 9    & 10   \\ \hline
				$d_n$ & 2    & 2    & 4    & 4    & 3    & 3    & 4    & 4    & 2    & 2    \\
				$L_n$ & 4    & 4    & 16    & 16    & 8    & 8    & 16    & 16    & 4    & 4    \\
				RMSE (m) & 1.29 & 1.3 & 0.99 & 1 & 1.12 & 1.12 & 1 & 1.01 & 1.29 & 1.31 \\
				Time (s) & 0.12 & 0.11 & 0.59 & 0.58 & 0.24 & 0.24 & 0.58 & 0.59 & 0.11 & 0.1 \\\hline
		\end{tabular}}
		\label{tab:9}
	\end{table}
	We evaluate the performance of different nodes in the WSN under the assumption that the measurement noise follows the distribution described in (\ref{alpha noise}). Table \ref{tab:9} presents the RMSE and simulation time of different nodes in the WSN (depicted in Fig. \ref{mao3}). Analysis of Table \ref{tab:9} reveals a positive correlation between the number of neighbors a node has and the accuracy of its estimation. As the number of neighbors increases, the MFDKF needs to fuse more measurement sub-models, resulting in an increased computational complexity.
	
\begin{figure}[!ht]
		\centering
		\includegraphics[width=0.9\linewidth]{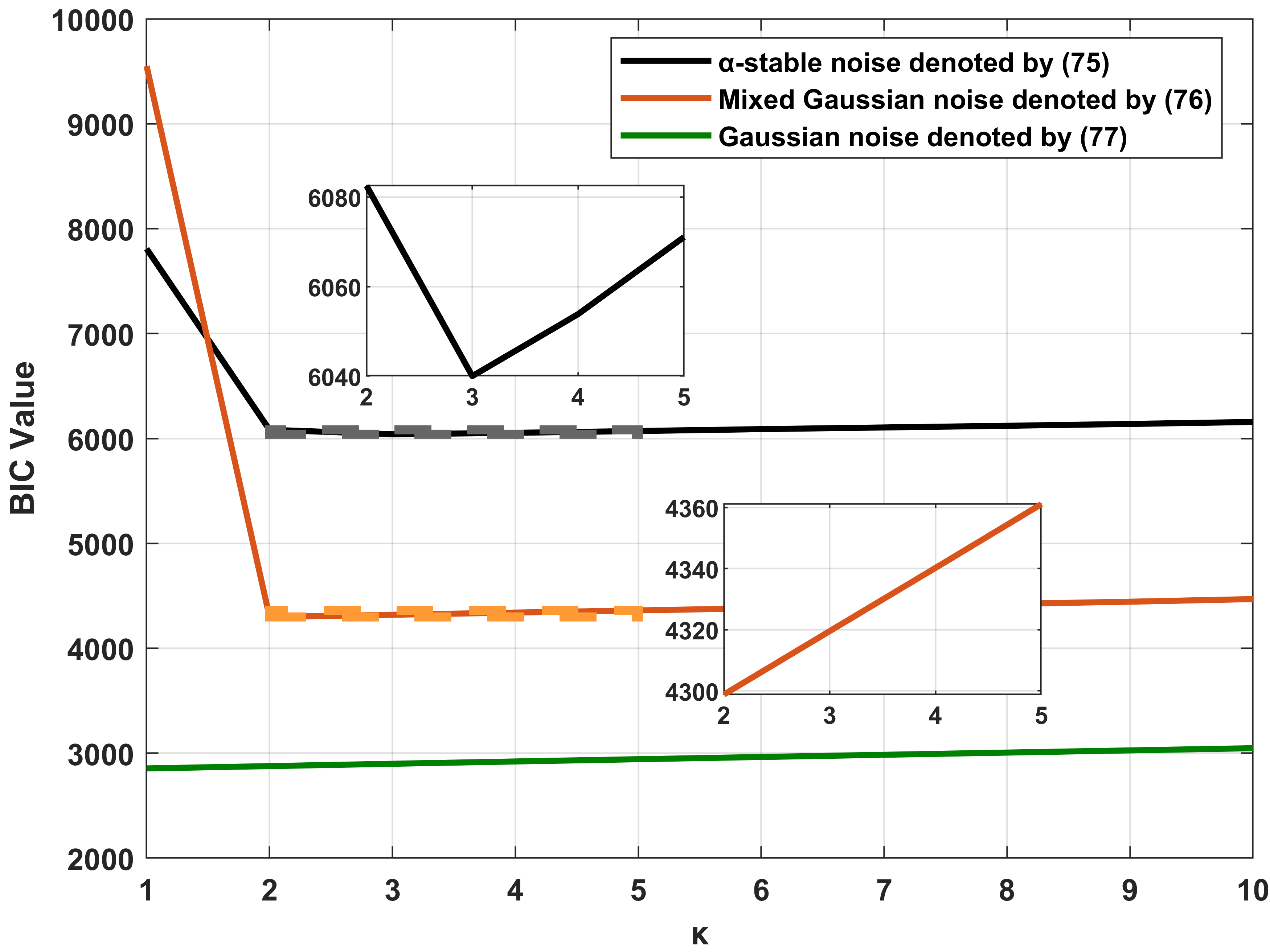} 
		\caption{The value of BIC with different values of $\kappa$ under different noises.}
		\label{mao7}
	\end{figure}

	\begin{table}[!ht]
		\caption{RMSEs of the MFDKF with different values of $\kappa_n$}
		\centering
		\begin{tabular}{llllllllll}
			\hline
			$\kappa_n$ & 2     & 3     & 4    & 5    \\ \hline
			$L_n$    & 16    & 81    & 256  & 625  \\
			RMSE (m)     & 1 & 0.85 & 0.81 & 0.76  \\
			Time (s)     & 0.57 & 8 & 64.6 & 381.54  \\ \hline
		\end{tabular}
		\label{tab:5}
	\end{table}
	In the same non-Gaussian environment, we compare the performance of MFDKF with different values of parameter $\kappa_n$. The performance of MFDKF improves with increasing values of $\kappa_n$, as indicated in Table \ref{tab:5}. This improvement can be attributed to the enhanced noise estimation achieved by increasing $\kappa_n$ through the EM algorithm. Essentially, the increased value of $\kappa_n$ allows for a more accurate approximation of the fusion measurement model by the measurement sub-models. As $\kappa_n$ increases, the performance improvement of the MFDKF diminishes, and the algorithm fails to converge when $\kappa_n = 6$. This is because an excessively large number of Gaussian components $\kappa_n$ can lead to overfitting when modeling non-Gaussian noise. To address this, the Bayesian Information Criterion (BIC) \cite{865189} is used to determine the optimal number of components in a GMM by balancing goodness-of-fit and model complexity. As shown in Fig. \ref{mao7}, the optimal value of \(\kappa_n\) is determined when the BIC value reaches its minimum or begins to plateau, suggesting that adding more components offers negligible improvement and risks overfitting.
	
	For varying topologies and applications, computational complexity must be considered. Table \ref{tab:5} shows that computation time increases significantly with larger $\kappa_n$, consistent with Section \ref{computational}. Therefore, $\kappa_n$ should be chosen by balancing performance requirements and computational cost.
	
	\subsection{Simulation of C-MFDKF and S-MFDKF} \label{simulation 2}
	The simulations above indicate the improvement of the proposed MFDKF. In this section, we examine the performance of the derived algorithms based on specific scenarios, using $\kappa_n=2$ as an illustrative example. It should be noted that, according to the convergence condition of the algorithm provided in Section \ref{derived MFDKF}, the parameter $\xi$ must take values within the range $(0,1)$. We quantify the disagreement among nodes in the WSN using  
	\begin{equation}
		\begin{array}{l}
			{\delta _k}(m) = \sqrt {\sum\limits_{n = 1}^N {{{\left\| {{{{\boldsymbol{\hat x}}}_{n,k}} - {{\boldsymbol{\mu }}_k}} \right\|}^2}} } ,
			{{\boldsymbol{\mu }}_k} = \frac{1}{N}\sum\limits_{n = 1}^N {{{{\boldsymbol{\hat x}}}_{n,k}}} ,
		\end{array}
	\end{equation}
	with $N$ denotes the total number of nodes. This metric reflects the consensus among all nodes in the WSN, with smaller values indicating a higher level of result consistency among nodes. To highlight the differences in algorithm consistency, we use decibels as the unit in the figures:
		\begin{equation}
			{\delta _k}\left( {dB} \right) = 20\log \left( {{{{\delta _k}\left( m \right)} \mathord{\left/
						{\vphantom {{{\delta _k}\left( m \right)} {1\left( m \right)}}} \right.
						\kern-\nulldelimiterspace} {1\left( m \right)}}} \right).
	\end{equation}
	\begin{figure}[!ht]
		\centering
		\includegraphics[width=0.9\linewidth]{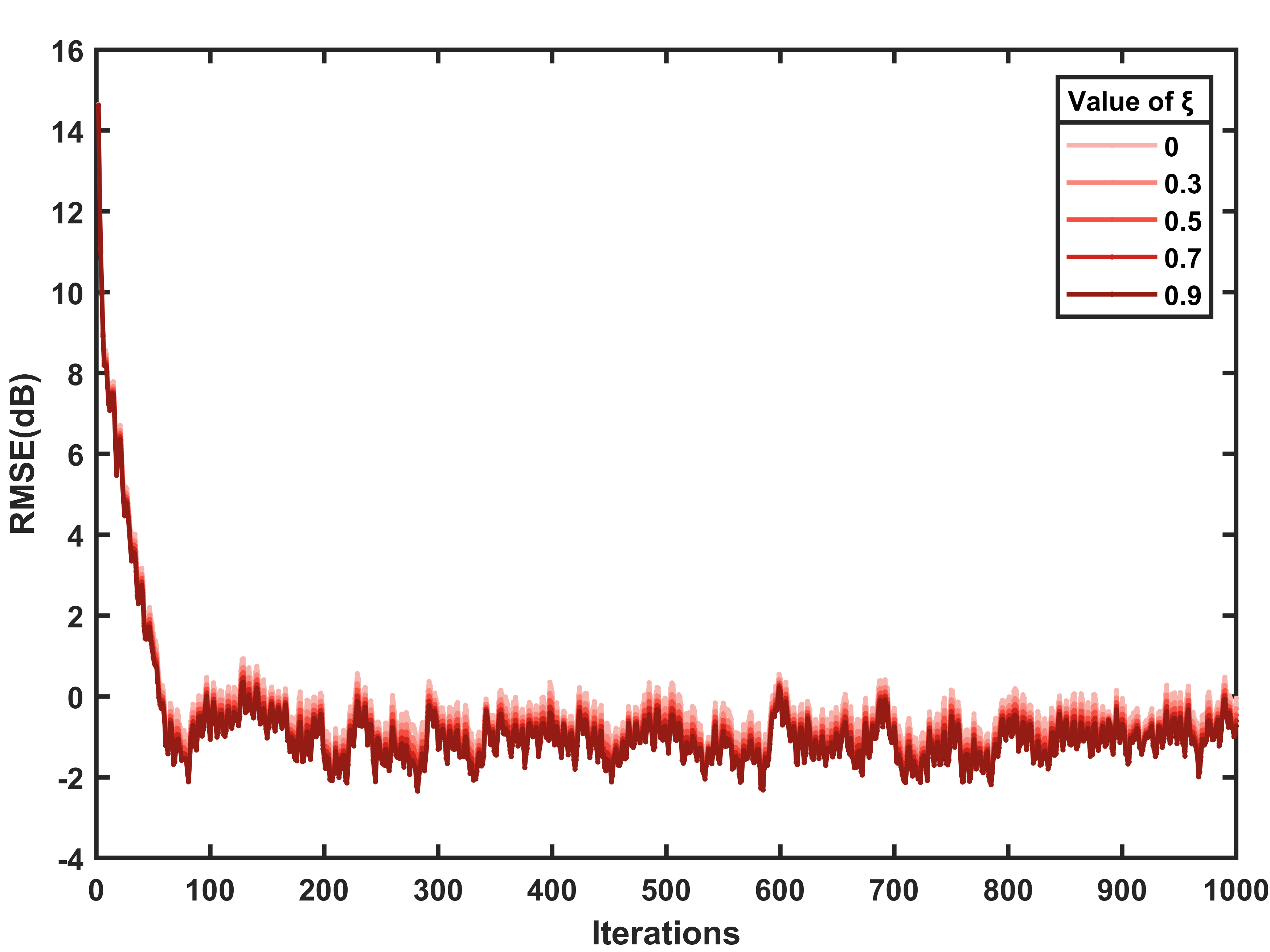} 
		\caption{Comparison of RMSEs of C-MFDKF with different values of $\xi$ under $\alpha$-stable noises, as represented by (\ref{alpha noise}).}
		\label{mao8}
	\end{figure}
	\begin{figure}[!ht]
		\centering
		\includegraphics[width=0.9\linewidth]{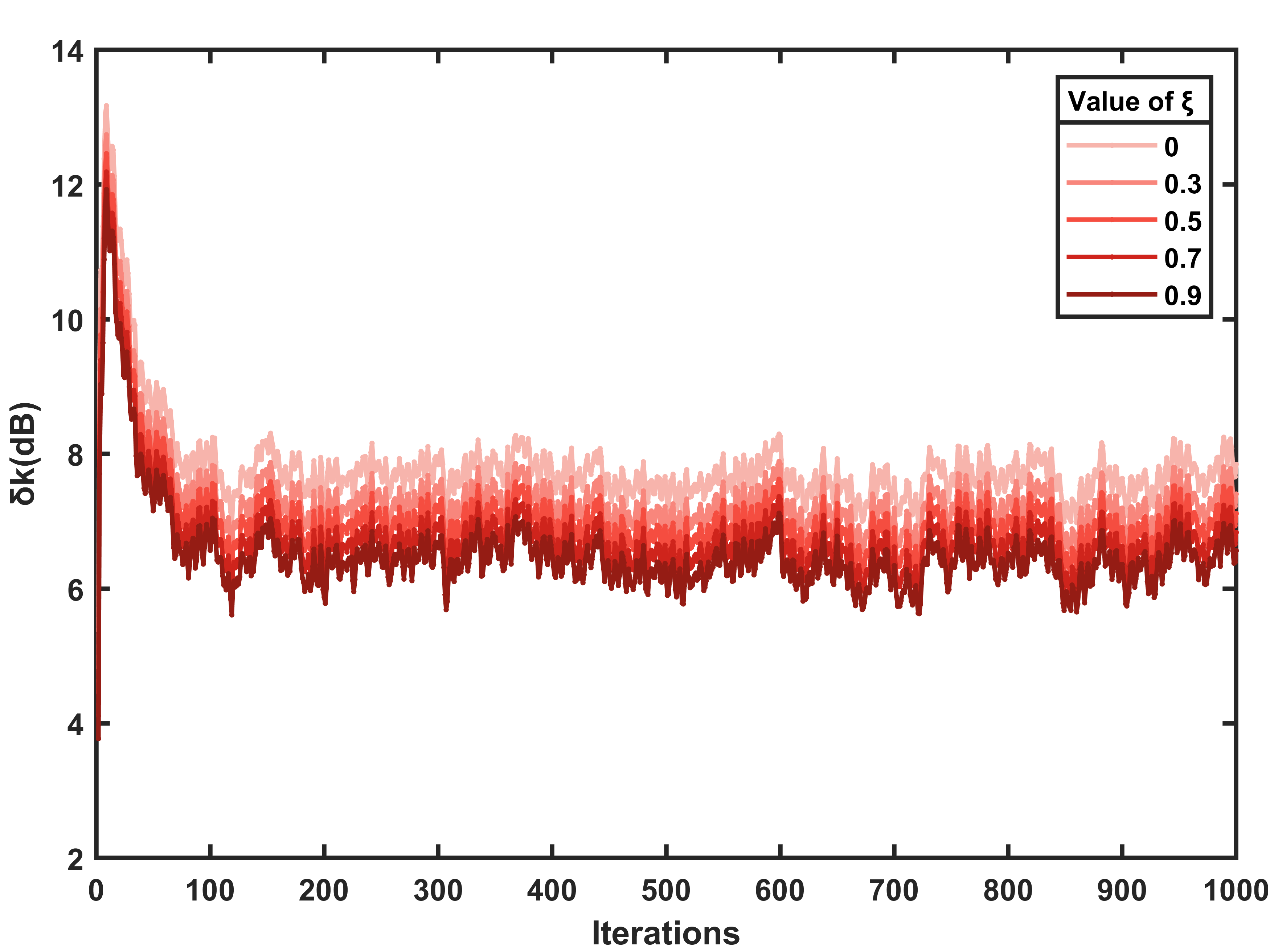}
		\caption{Comparison of disagreements of C-MFDKF with different $\xi$ under $\alpha$-stable noises, as represented by (\ref{alpha noise}).}
		\label{mao9}
	\end{figure}
	
	In the first scenario, we examine the WSN with a strong requirement for consensus among sensor nodes. We assume unrestricted communication between neighboring nodes, enabling the fulfillment of all information interactions in C-MFDKF. Consider the linear system discussed in Section \ref{simulation 1} as an illustrative example, where the measurement noises are modeled as $\alpha$-stable noise, as denoted in (\ref{alpha noise}). The RMSEs and disagreements of the C-MFDKF with different values of $\xi$ are shown in Fig.  \ref{mao8} and Fig. \ref{mao9}. Based on the observed performances, we can conclude that C-MFDKF successfully mitigates the disagreement between nodes while guaranteeing accurate estimation. Increasing $\xi$ within the specified range $(0,1)$ enhances the C-MFDKF's effectiveness in minimizing disagreement between nodes. The details of the simulation are shown in Table \ref{tab:8}. The performance of C-MFDKF ($\xi = 0.9$) under different non-Gaussian noise is shown in Table \ref{tab:4}, which indicates the RMSE of C-MFDKF is better. The simulation time (listed in Table \ref{Simulation time}) of C-MFDKF is longer than that of MFDKF, which corroborates the analysis of computational complexity in (\ref{computation_cmfdkf}).
	\begin{table}[!ht]
		\caption{RMSEs and disagreements of C-MFDKF with different $\xi$}
		\centering
			\begin{tabular}{cccccc}
				\hline
				$\xi$           & 0    & 0.3  & 0.5 & 0.7 & 0.9 \\ \hline
				RMSE (m)         & 1 & 0.97 & 0.96  & 0.94 & 0.93 \\
				Disagreement (m) & 2.47 & 2.35 & 2.27 & 2.19 & 2.13 \\ \hline
		\end{tabular}
		\label{tab:8}
	\end{table}
	
	Consider a scenario where information exchange between adjacent nodes in the WSN is limited. It is assumed that neighboring nodes are only permitted to exchange LKF results. Under these circumstances, the MFDKF cannot function as it requires measurements from neighboring nodes. The performances of the S-MFDKF with different parameters $\xi$ are compared. Table \ref{tab:10} shows that both the accuracy and the consensus are improved compared to $\xi=0$. Increasing $\xi$ within the specified range $(0,1)$ enhances the S-MFDKF's capability to reduce both the estimation error and disagreement among nodes.
	\begin{table}[!ht]
		\caption{RMSEs and disagreements of S-MFDKF with different $\xi$}
		\centering
			\begin{tabular}{cccccc}
				\hline
				$\xi$           & 0    & 0.3  & 0.5 & 0.7 & 0.9 \\ \hline
				RMSE (m)         & 1.67 & 1.62 & 1.57  & 1.52 & 1.48 \\
				Disagreement (m) & 4.1 & 3.77 & 3.56 & 3.37 & 3.19 \\ \hline
		\end{tabular}
		\label{tab:10}
	\end{table}
	
	The performance of S-MFDKF ($\xi = 0.9$) under different non-Gaussian noise is shown in Table \ref{tab:4}. It can be concluded that although S-MFDKF is less effective than MFDKF, it is still better compared to DMCKF and DMEEKF. The simulation time (listed in Table \ref{Simulation time}) of S-MFDKF is shorter than that of MFDKF, which corroborates analysis in (\ref{computation_smfdkf}). Moreover, it is faster than DMEEKF.
	
	\section{Conclusion} \label{conclusions}
	This work proposes an MFDKF to address the issue of non-Gaussian measurement noise in WSNs. The MFDKF employs a GMM to approximate the noise, allowing the measurement model of the DKF to be expressed as a combination of sub-models with varying Gaussian components. The final estimation is obtained by fusing all sub-models according to the probability of each model's occurrence. Considering the high consensus requirements and limited communication in the WSN, we introduce two derivative algorithms: C-MFDKF and S-MFDKF. Mean error analysis and mean-square error analysis were conducted to validate the convergence of the proposed MFDKF and its derived algorithms. Simulations confirmed the favorable performance of the proposed MFDKF and its derived algorithms. Future work will focus on reducing model numbers and extending the algorithm to nonlinear systems.
	
	\bibliographystyle{IEEEtran}
	\bibliography{mybibfile}

\begin{thebibliography}{10}
\providecommand{\url}[1]{#1}
\csname url@samestyle\endcsname
\providecommand{\newblock}{\relax}
\providecommand{\bibinfo}[2]{#2}
\providecommand{\BIBentrySTDinterwordspacing}{\spaceskip=0pt\relax}
\providecommand{\BIBentryALTinterwordstretchfactor}{4}
\providecommand{\BIBentryALTinterwordspacing}{\spaceskip=\fontdimen2\font plus
\BIBentryALTinterwordstretchfactor\fontdimen3\font minus
  \fontdimen4\font\relax}
\providecommand{\BIBforeignlanguage}[2]{{%
\expandafter\ifx\csname l@#1\endcsname\relax
\typeout{** WARNING: IEEEtran.bst: No hyphenation pattern has been}%
\typeout{** loaded for the language `#1'. Using the pattern for}%
\typeout{** the default language instead.}%
\else
\language=\csname l@#1\endcsname
\fi
#2}}
\providecommand{\BIBdecl}{\relax}
\BIBdecl

\bibitem{ma2022method}
M.~Ma, B.~He, N.~Wang, and R.~Shen, ``A method for monitoring the solar
  resources of high-scale photovoltaic power plants based on wireless sensor
  networks,'' \emph{Sustainable Energy Technologies and Assessments}, vol.~53,
  p. 102678, 2022.

\bibitem{zhang2022consensus}
C.~Zhang, J.~Qin, H.~Li, Y.~Wang, S.~Wang, and W.~X. Zheng, ``Consensus-based
  distributed two-target tracking over wireless sensor networks,''
  \emph{Automatica}, vol. 146, p. 110593, 2022.

\bibitem{liu2020multistep}
F.~Liu, C.~Jiang, and W.~Xiao, ``Multistep prediction-based adaptive dynamic
  programming sensor scheduling approach for collaborative target tracking in
  energy harvesting wireless sensor networks,'' \emph{IEEE Transactions on
  Automation Science and Engineering}, vol.~18, no.~2, pp. 693--704, 2020.

\bibitem{talebi2016distributed}
S.~P. Talebi, S.~Kanna, and D.~P. Mandic, ``A distributed quaternion kalman
  filter with applications to smart grid and target tracking,'' \emph{IEEE
  Transactions on Signal and Information Processing over Networks}, vol.~2,
  no.~4, pp. 477--488, 2016.

\bibitem{YANG20142070}
W.~Yang, G.~Chen, X.~Wang, and L.~Shi, ``Stochastic sensor activation for
  distributed state estimation over a sensor network,'' \emph{Automatica},
  vol.~50, no.~8, pp. 2070--2076, 2014.

\bibitem{he2020distributed}
S.~He, H.-S. Shin, S.~Xu, and A.~Tsourdos, ``Distributed estimation over a
  low-cost sensor network: A review of state-of-the-art,'' \emph{Information
  Fusion}, vol.~54, pp. 21--43, 2020.

\bibitem{sun2017multi}
S.~Sun, H.~Lin, J.~Ma, and X.~Li, ``Multi-sensor distributed fusion estimation
  with applications in networked systems: A review paper,'' \emph{Information
  Fusion}, vol.~38, pp. 122--134, 2017.

\bibitem{sun2004multi}
S.-L. Sun and Z.-L. Deng, ``Multi-sensor optimal information fusion kalman
  filter,'' \emph{Automatica}, vol.~40, no.~6, pp. 1017--1023, 2004.

\bibitem{modalavalasa2021review}
S.~Modalavalasa, U.~K. Sahoo, A.~K. Sahoo, and S.~Baraha, ``A review of robust
  distributed estimation strategies over wireless sensor networks,''
  \emph{Signal Processing}, vol. 188, p. 108150, 2021.

\bibitem{olfati2005distributed}
R.~Olfati-Saber, ``Distributed kalman filter with embedded consensus filters,''
  in \emph{Proceedings of the 44th IEEE Conference on Decision and
  Control}.\hskip 1em plus 0.5em minus 0.4em\relax IEEE, 2005, pp. 8179--8184.

\bibitem{olfati2007consensus}
R.~Olfati-Saber, J.~A. Fax, and R.~M. Murray, ``Consensus and cooperation in
  networked multi-agent systems,'' \emph{Proceedings of the IEEE}, vol.~95,
  no.~1, pp. 215--233, 2007.

\bibitem{olfati2007distributed}
R.~Olfati-Saber, ``Distributed kalman filtering for sensor networks,'' in
  \emph{2007 46th IEEE Conference on Decision and Control}.\hskip 1em plus
  0.5em minus 0.4em\relax IEEE, 2007, pp. 5492--5498.

\bibitem{olfati2009kalman}
------, ``Kalman-consensus filter: Optimality, stability, and performance,'' in
  \emph{Proceedings of the 48h IEEE Conference on Decision and Control (CDC)
  held jointly with 2009 28th Chinese Control Conference}.\hskip 1em plus 0.5em
  minus 0.4em\relax Ieee, 2009, pp. 7036--7042.

\bibitem{he2022generalized}
J.~He, G.~Wang, H.~Yu, J.~Liu, and B.~Peng, ``Generalized minimum error entropy
  kalman filter for non-gaussian noise,'' \emph{ISA transactions}, 2022.

\bibitem{wang2022centralized}
G.~Wang, C.~Yang, L.~Ma, and W.~Dai, ``Centralized and distributed robust state
  estimation over sensor networks using elliptical distribution,'' \emph{IEEE
  Internet of Things Journal}, vol.~9, no.~21, pp. 21\,825--21\,837, 2022.

\bibitem{he2024gaussian}
J.~He, B.~Peng, Z.~Feng, S.~Zhong, B.~He, and G.~Wang, ``A gaussian mixture
  unscented rauch--tung--striebel smoothing framework for trajectory
  reconstruction,'' \emph{IEEE Transactions on Industrial Informatics}, 2024.

\bibitem{9167469}
M.~Bai, Y.~Huang, Y.~Zhang, and F.~Chen, ``A novel heavy-tailed mixture
  distribution based robust kalman filter for cooperative localization,''
  \emph{IEEE Transactions on Industrial Informatics}, vol.~17, no.~5, pp.
  3671--3681, 2021.

\bibitem{fan2022background}
X.~Fan, G.~Wang, J.~Han, and Y.~Wang, ``A background-impulse kalman filter with
  non-gaussian measurement noises,'' \emph{IEEE Transactions on Systems, Man,
  and Cybernetics: Systems}, 2022.

\bibitem{coates2004distributed}
M.~Coates, ``Distributed particle filters for sensor networks,'' in
  \emph{Proceedings of the 3rd international symposium on Information
  processing in sensor networks}, 2004, pp. 99--107.

\bibitem{yu2016distributed}
J.~Y. Yu, M.~J. Coates, M.~G. Rabbat, and S.~Blouin, ``A distributed particle
  filter for bearings-only tracking on spherical surfaces,'' \emph{IEEE Signal
  Processing Letters}, vol.~23, no.~3, pp. 326--330, 2016.

\bibitem{li2017distributed}
J.~Li and A.~Nehorai, ``Distributed particle filtering via optimal fusion of
  gaussian mixtures,'' \emph{IEEE Transactions on Signal and Information
  Processing over Networks}, vol.~4, no.~2, pp. 280--292, 2017.

\bibitem{xu2018distributed}
C.~Xu, S.~Zhao, B.~Huang, and F.~Liu, ``Distributed student's t filtering
  algorithm for heavy-tailed noises,'' \emph{International Journal of Adaptive
  Control and Signal Processing}, vol.~32, no.~6, pp. 875--890, 2018.

\bibitem{yan2020distributed}
L.~Yan, C.~Di, Q.~J. Wu, Y.~Xia, and S.~Liu, ``Distributed fusion estimation
  for multisensor systems with non-gaussian but heavy-tailed noises,''
  \emph{ISA transactions}, vol. 101, pp. 160--169, 2020.

\bibitem{li2022multi}
T.~Li, Z.~Hu, Z.~Liu, and X.~Wang, ``Multi-sensor suboptimal fusion student's $
  t $ filter,'' \emph{IEEE Transactions on Aerospace and Electronic Systems},
  2022.

\bibitem{8214971}
Y.~Huang, Y.~Zhang, P.~Shi, Z.~Wu, J.~Qian, and J.~A. Chambers, ``Robust kalman
  filters based on gaussian scale mixture distributions with application to
  target tracking,'' \emph{IEEE Transactions on Systems, Man, and Cybernetics:
  Systems}, vol.~49, no.~10, pp. 2082--2096, 2019.

\bibitem{chen2017maximum}
B.~Chen, X.~Liu, H.~Zhao, and J.~C. Principe, ``Maximum correntropy kalman
  filter,'' \emph{Automatica}, vol.~76, pp. 70--77, 2017.

\bibitem{chen2019minimum}
B.~Chen, L.~Dang, Y.~Gu, N.~Zheng, and J.~C. Pr{\'\i}ncipe, ``Minimum error
  entropy kalman filter,'' \emph{IEEE Transactions on Systems, Man, and
  Cybernetics: Systems}, vol.~51, no.~9, pp. 5819--5829, 2019.

\bibitem{wang2021numerically}
G.~Wang, B.~Chen, X.~Yang, B.~Peng, and Z.~Feng, ``Numerically stable minimum
  error entropy kalman filter,'' \emph{Signal Processing}, vol. 181, p. 107914,
  2021.

\bibitem{he2023generalized}
J.~He, G.~Wang, K.~Cao, H.~Diao, G.~Wang, and B.~Peng, ``Generalized minimum
  error entropy for robust learning,'' \emph{Pattern Recognition}, vol. 135, p.
  109188, 2023.

\bibitem{wang2019distributed}
G.~Wang, R.~Xue, and J.~Wang, ``A distributed maximum correntropy kalman
  filter,'' \emph{Signal Processing}, vol. 160, pp. 247--251, 2019.

\bibitem{liu2007correntropy}
W.~Liu, P.~P. Pokharel, and J.~C. Principe, ``Correntropy: Properties and
  applications in non-gaussian signal processing,'' \emph{IEEE Transactions on
  signal processing}, vol.~55, no.~11, pp. 5286--5298, 2007.

\bibitem{wang2021distributed}
G.~Wang, N.~Li, and Y.~Zhang, ``Distributed maximum correntropy linear and
  nonlinear filters for systems with non-gaussian noises,'' \emph{Signal
  Processing}, vol. 182, p. 107937, 2021.

\bibitem{hu2022efficient}
C.~Hu and B.~Chen, ``An efficient distributed kalman filter over sensor
  networks with maximum correntropy criterion,'' \emph{IEEE Transactions on
  Signal and Information Processing over Networks}, vol.~8, pp. 433--444, 2022.

\bibitem{zhang2015convergence}
Y.~Zhang, B.~Chen, X.~Liu, Z.~Yuan, and J.~C. Principe, ``Convergence of a
  fixed-point minimum error entropy algorithm,'' \emph{Entropy}, vol.~17,
  no.~8, pp. 5549--5560, 2015.

\bibitem{8267224}
A.~R. Heravi and G.~A. Hodtani, ``A new information theoretic relation between
  minimum error entropy and maximum correntropy,'' \emph{IEEE Signal Processing
  Letters}, vol.~25, no.~7, pp. 921--925, 2018.

\bibitem{feng2023distributed}
Z.~Feng, G.~Wang, B.~Peng, J.~He, and K.~Zhang, ``Distributed minimum error
  entropy kalman filter,'' \emph{Information Fusion}, vol.~91, pp. 556--565,
  2023.

\bibitem{reynolds2009gaussian}
D.~A. Reynolds, ``Gaussian mixture models.'' \emph{Encyclopedia of biometrics},
  vol. 741, no. 659-663, 2009.

\bibitem{fan2021interacting}
X.~Fan, G.~Wang, J.~Han, and Y.~Wang, ``Interacting multiple model based on
  maximum correntropy kalman filter,'' \emph{IEEE Transactions on Circuits and
  Systems II: Express Briefs}, vol.~68, no.~8, pp. 3017--3021, 2021.

\bibitem{9661360}
S.~Chen, Q.~Zhang, T.~Zhang, L.~Zhang, L.~Peng, and S.~Wang, ``Robust state
  estimation with maximum correntropy rotating geometric unscented kalman
  filter,'' \emph{IEEE Transactions on Instrumentation and Measurement},
  vol.~71, pp. 1--14, 2022.

\bibitem{min2019robust}
Z.~Min, J.~Wang, and M.~Q.-H. Meng, ``Robust generalized point cloud
  registration with orientational data based on expectation maximization,''
  \emph{IEEE Transactions on Automation Science and Engineering}, vol.~17,
  no.~1, pp. 207--221, 2019.

\bibitem{guo2017augmented}
F.~Guo, O.~Wu, H.~Kodamana, Y.~Ding, and B.~Huang, ``An augmented model
  approach for identification of nonlinear errors-in-variables systems using
  the em algorithm,'' \emph{IEEE Transactions on Systems, Man, and Cybernetics:
  Systems}, vol.~48, no.~11, pp. 1968--1978, 2017.

\bibitem{9655313}
Y.~Luo, J.~Zhou, and W.~Yang, ``Distributed state estimation with colored
  noises,'' \emph{IEEE Transactions on Circuits and Systems II: Express
  Briefs}, vol.~69, no.~6, pp. 2807--2811, 2022.

\bibitem{xie2019adaptive}
G.~Xie, L.~Sun, T.~Wen, X.~Hei, and F.~Qian, ``Adaptive transition probability
  matrix-based parallel imm algorithm,'' \emph{IEEE Transactions on Systems,
  Man, and Cybernetics: Systems}, vol.~51, no.~5, pp. 2980--2989, 2019.

\bibitem{ZHONG2024111343}
S.~Zhong, B.~Peng, J.~He, Z.~Feng, M.~Li, and G.~Wang, ``Kalman filtering based
  on dynamic perception of measurement noise,'' \emph{Mechanical Systems and
  Signal Processing}, vol. 213, p. 111343, 2024.

\bibitem{CHEN2019228}
B.~Chen, G.~Hu, D.~W. Ho, and L.~Yu, ``Distributed kalman filtering for
  time-varying discrete sequential systems,'' \emph{Automatica}, vol.~99, pp.
  228--236, 2019.

\bibitem{865189}
C.~Biernacki, G.~Celeux, and G.~Govaert, ``Assessing a mixture model for
  clustering with the integrated completed likelihood,'' \emph{IEEE
  Transactions on Pattern Analysis and Machine Intelligence}, vol.~22, no.~7,
  pp. 719--725, 2000.

\end{thebibliography}
	\begin{IEEEbiography}
		[{\includegraphics[width=1in,height=1.25in,clip,keepaspectratio]{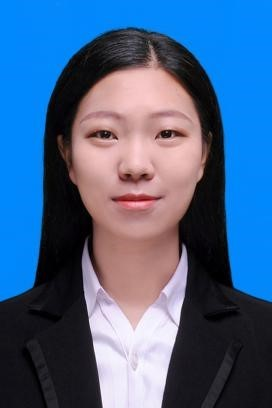}}] 
		Xuemei Mao received a B.S. degree in mechanical design, manufacturing, and automation from Xidian University, Xian, China, in 2020. She is currently pursuing a Ph.D. degree in mechanical engineering with the School of Mechanical and Electrical Engineering, University of Electronic Science and Technology of China, Chengdu, China. Her current research interests include signal processing, distributed systems, and target tracking.
	\end{IEEEbiography}
	\begin{IEEEbiography}
		[{\includegraphics[width=1in,height=1.25in,clip,keepaspectratio]{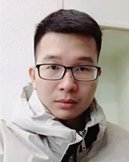}}] 
		Jiacheng received a B.S. degree in mechanical engineering from the University of Electronic Science and Technology of China, Chengdu, China, in 2020. He is currently pursuing a Ph.D. degree in the School of Mechanical and Electrical Engineering, University of Electronic Science and Technology of China, Chengdu, China. His current research interests include information-theoretic learning, signal processing, and adaptive filtering.
	\end{IEEEbiography}
	\begin{IEEEbiography}
		[{\includegraphics[width=1in,height=1.25in,clip,keepaspectratio]{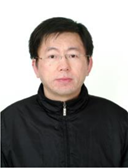}}] 
		Gang Wang received a B.E. degree in Communication Engineering and a Ph.D. degree in Biomedical Engineering from the University of Electronic Science and Technology of China, Chengdu, China, in 1999 and 2008, respectively. In 2009, he joined the School of Information and Communication Engineering, University of Electronic Science and Technology of China, China, where he is currently an Associate Professor. His current research interests include signal processing and intelligent systems.
	\end{IEEEbiography}
	\begin{IEEEbiography}
		[{\includegraphics[width=1in,height=1.25in,clip,keepaspectratio]{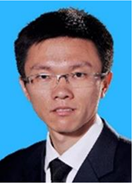}}] 
		Bei Peng received a B.S. degree in mechanical engineering from Beihang University, Beijing, China, in 1999, and the M.S. and Ph.D. degrees in mechanical engineering from Northwestern University, Evanston, IL, USA, in 2003 and 2008, respectively. He is currently a Full Professor of Mechanical Engineering at the University of Electronic Science and Technology of China, Chengdu, China. He holds 30 authorized patents. He has served as a PI or a CoPI for more than ten research projects, including the National Science Foundation of China. His research interests mainly include intelligent manufacturing systems, robotics, and its applications.
	\end{IEEEbiography}
	\begin{IEEEbiography}
		[{\includegraphics[width=1in,height=1.25in,clip,keepaspectratio]{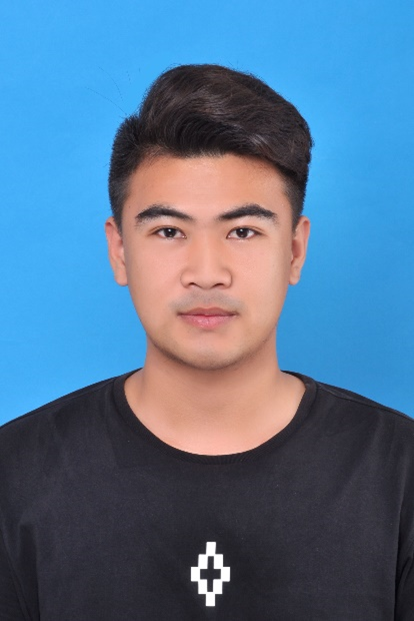}}] 
		Kun Zhang received a B.S. degree in electronic and information engineering from Hainan University, Haikou, China, in 2018. He is currently working toward a Ph.D. degree in mechanical engineering with the Mechanical Engineering of the University of Electronic Science and Technology of China, Chengdu, China. His research interests include intelligent manufacturing systems, robotics, and its applications.
	\end{IEEEbiography}
	\begin{IEEEbiography}
		[{\includegraphics[width=1in,height=1.25in,clip,keepaspectratio]{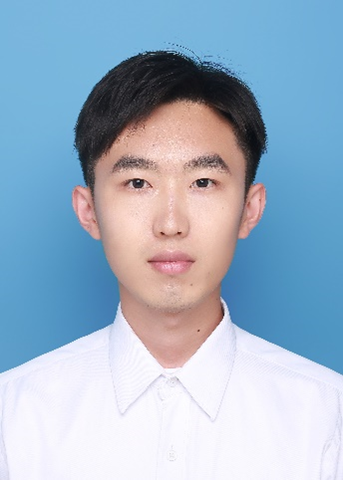}}] 
		Song Gao received a B.S. degree in mechanical design, manufacturing, and automation from the University of Electronic Science and Technology of China (UESTC), in 2020. He is currently pursuing a Ph.D. degree in mechanical engineering with the School of Mechanical and Electrical Engineering of UESTC, Chengdu, China. His current research interests include signal processing, multi-agent systems, and consensus control.
	\end{IEEEbiography}
	\begin{IEEEbiography}
		[{\includegraphics[width=1in,height=1.25in,clip,keepaspectratio]{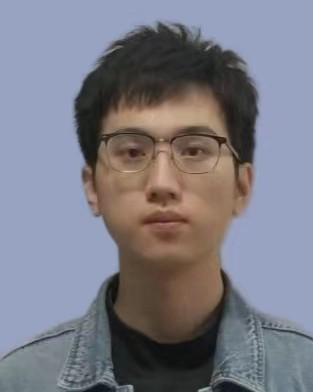}}] 
		Jian Chen received a B.S. degree in mechanical engineering from Beihang University, Beijing, China, in 2020. He is currently pursuing a Ph.D. degree in the School of Mechanical and Electrical Engineering, University of Electronic Science and Technology of China, Chengdu, China. His current research interests include machine learning and signal processing.
	\end{IEEEbiography}
	\vfill
\end{document}